%% file: sparse_X_manuscript.tex
\documentclass[journal]{IEEEtran}

\def\BibTeX{{\rm B\kern-.05em{\sc i\kern-.025em b}\kern-.08em
    T\kern-.1667em\lower.7ex\hbox{E}\kern-.125emX}}

\include{def}

\usepackage{bm}

\usepackage{subfig}

\usepackage{color}
\usepackage{graphics}
\usepackage{graphicx}
\usepackage{caption}
\usepackage{tabularx}
\usepackage{epstopdf}

\usepackage{algpseudocode}
\usepackage[ruled,linesnumbered]{algorithm2e}

\usepackage{xr}
\usepackage{xr-hyper}
\def\diag{{\rm diag}}
\date{}
\begin{document}
\title{Sparse Range-Doppler Profile Recovery in Colored Interference for Adaptive Radar}
\author{Tao Wang, Augusto Aubry, Antonio De Maio, Junli Liang
\thanks{Tao Wang and Junli Liang are with School of Electronics and Information, Northwestern Polytechnical University, Xi'an, China. (e-mail: wang.tao@mail.nwpu.edu.cn; liangjunli@nwpu.edu.cn ); Augusto Aubry and Antonio De Maio are with the Department of Electrical Engineering and Information Technology (DIETI), Università degli Studi di Napoli "Federico II," 80125 Naples, Italy, and also with the National Inter-University Consortium for Telecommunications, 43124 Parma, Italy (e-mail: augusto.aubry@unina.it; ademaio@unina.it). \textit{(Corresponding author: Augusto Aubry.)}
}}
\markboth{IEEE Transactions on Signal Processing}{Shell \MakeLowercase{\textit{et al.}}: Bare Demo of
IEEEtran.cls for Journals}
\input{epsf}

\maketitle
\input{epsf}
\begin{sloppypar}
\begin{abstract}
	Range-Doppler profile (RDP) recovery is a fundamental task in adaptive radar signal processing, whose performance can be severely degraded by Doppler grid mismatch and colored interference. This article develops a unified sparsity-aware framework for on-grid and off-grid RDP recovery in the presence of \textcolor{black}{correlated} clutter with unknown \textcolor{black}{spectral properties}. The problem is formulated within a regularized maximum likelihood framework by exploiting both primary and secondary data, where the sparse RDP and interference covariance are jointly inferred, while Doppler frequency displacements are additionally estimated in the off-grid case to mitigate basis mismatch. Both covariance-adaptive one-step formulations and their sample-covariance-matrix-based two-step alternatives are developed within the \textcolor{black}{mentioned} framework. The resulting challenging nonconvex optimization problems
	are efficiently tackled by a majorization-minimization (MM)
	algorithm for the on-grid case and an MM-minimum block
	improvement (MBI) \textcolor{black}{approach} for the off-grid \textcolor{black}{counterpart}. Furthermore,
	Bayesian information criterion (BIC)-based refinement strategies
	are introduced to improve RDP recovery and Doppler frequency displacement estimation. 
	Numerical results under colored interference demonstrate the superior recovery performance of the proposed methods over representative benchmarks.
\end{abstract}


\section{Introduction}

\textcolor{black}{
Range-Doppler imaging is a fundamental component of adaptive radar signal processing, supporting localization, tracking \cite{9457493}, \textcolor{black}{recognition \cite{9673798,8695853},} and resource allocation \cite{10196038}.}
\textcolor{black}{Conventional RDPs are obtained by range matched filtering followed by fast Fourier transform (FFT)-based Doppler processing across coherent pulses \cite{7486028}. Although computationally efficient, this processing does not exploit the intrinsic sparsity of many practical range-Doppler scenes, where \textcolor{black}{target scatterers occupy only a small fraction of the state space \cite{7484756,8708966}. This observation has motivated sparsity-driven RDP retrieval methods to improve inference performance \cite{6850186,5960585}.}}

Existing sparse RDP \textcolor{black}{estimation} methods can be broadly categorized into on-grid and off-grid paradigms. On-grid methods presuppose that the target parameters exactly coincide with a predefined range-Doppler grid. Representative techniques include compressed sensing methods for incomplete \cite{7485322,8639010} or sub-Nyquist measurements \cite{8598758}, iterative adaptive approaches \cite{5417172}, and sparse Bayesian learning methods \cite{1315936,4244754,5342501}. 
However, most of these approaches are developed under simplified noise models and do not explicitly account for correlated interference. To address this issue, several studies have extended sparse on-grid RDP recovery to correlated clutter and colored-interference environments. Within a modified smoothed-$l_0$ framework, \cite{9447912} exploits two-dimensional block sparsity for weak-target detection in clutter. Moreover, \cite{8675515} and \cite{8835657} incorporate interference statistical characteristics into sparse RDP inference. However, \cite{8675515} constrains the interference covariance to a dictionary-dependent structure, whereas \cite{8835657} considers colored Gaussian interference-plus-noise. Despite their enhanced capability to handle correlated disturbances, these methods remain restricted to a predefined grid and are susceptible to basis mismatch \textcolor{black}{impairments}.

To alleviate the basis mismatch inherent in on-grid modeling, off-grid sparse RDP recovery methods refine the discretized dictionary or directly estimate continuous parameter deviations. In \cite{teke2014robust}, a parameter-perturbed orthogonal matching pursuit (OMP) method iteratively adjusts the selected atoms for \textcolor{black}{the} off-grid RDP reconstruction under additive noise. Similarly, \cite{sun2020range} adopts a coarse-to-fine matched-filter OMP strategy, where a coarse range-Doppler cell is first identified and then locally refined on a denser grid. Beyond dictionary refinement, other approaches explicitly parameterize the off-grid mismatch and jointly estimate the sparse coefficients and continuous offsets. The regularized maximum likelihood (RML) frameworks in \cite{10891687,11204963} jointly estimate sparse angle-frequency \textcolor{black}{profile}, noise power, and continuous angular offsets. 
\textcolor{black}{
Gridless methods avoid discretization by directly estimating continuous parameters, with atomic-norm approaches developed for frequency estimation \cite{6576276}, regular-array DOA estimation \cite{7314978}, and range-velocity estimation in joint radar-communication systems \cite{zheng2025gridless}.
}
Despite their improved robustness to basis mismatch, these methods are predominantly developed for additive-noise environments and generally do not explicitly account for colored interference.
	
\textcolor{black}{In contrast, fewer studies have considered off-grid RDP recovery in colored interference. In \cite{8387428}, a hierarchical Bayesian method jointly estimates the RDP, grid mismatch, and colored disturbance, but the disturbance covariance is confined to a finite-order autoregressive model. For passive bistatic radar, \cite{8241346} develops an atomic-norm-based gridless formulation for angle-range-Doppler estimation in clutter and noise, where \textcolor{black}{the} dominant interference is removed through a separate cancellation stage. Thus, existing methods generally rely on specific disturbance models or processing architectures, limiting their applicability to general colored-interference environments.}


\textcolor{black}{In summary, existing sparse RDP recovery methods either primarily assume white-noise-dominated environments or rely on structured disturbance models, limiting their applicability to general colored interference.}
Motivated by these limitations, this article develops a secondary-data-aided, sparsity-aware framework for on-grid and off-grid RDP recovery with an unknown, unstructured interference-plus-noise covariance matrix. Specifically, the problem is formulated within an RML framework, where the sparse RDP, covariance matrix \textcolor{black}{and possible Doppler frequency displacement} are jointly inferred from primary and secondary data \cite{10077421}. 


The resultant problems are nonconvex due to coupling among the RDP, covariance matrix, and possible Doppler frequency displacements. Tailored majorization-minimization (MM)-based algorithms are developed: a standard MM procedure is used for the on-grid case, while an MM-based minimum block improvement (MM-MBI) scheme \textcolor{black}{\cite{razaviyayn2013unified,8454321}} is adopted for the off-grid \textcolor{black}{situation} to refine the RDP and Doppler displacements. \textcolor{black}{Two-step schemes are also developed for both scenarios, where the covariance matrix is substituted with the sample covariance matrix (SCM).} Furthermore, a BIC-based refinement procedure is incorporated to identify dominant RDP components and suppress spurious responses, followed by support-conditioned re-estimation of the RDP, covariance matrix, and possible Doppler displacements.

The main technical contributions of this article can be streamlined as follows.
\begin{enumerate}
	\item A unified sparsity-aware framework is developed for on-grid and off-grid RDP recovery under general colored interference \textcolor{black}{scenario}. The one-step RML formulation jointly estimates the sparse RDP and the unknown, unstructured interference-plus-noise covariance matrix, together with Doppler frequency displacements in the off-grid case, while the two-step alternative employs the SCM estimated from secondary data \textcolor{black}{before regularized likelihood optimization}.
	\item 
	Tailored MM and MM-MBI algorithms are developed for the on-grid and off-grid problems, respectively, enabling joint estimation of the RDP, covariance matrix, and possible Doppler frequency displacements.
	\item 
	The convergence properties of the proposed algorithms are established, where the objective sequence is monotonically nonincreasing, and \textcolor{black}{any} cluster point satisfies the corresponding Karush-Kuhn-Tucker conditions.
	\item 
	Numerical results demonstrate that joint covariance estimation generally provides superior RDP recovery over the SCM-based alternatives and conventional methods, while the off-grid formulations effectively mitigate Doppler grid mismatch \textcolor{black}{issues}.
\end{enumerate}

The rest of this article is organized as follows. Section II introduces the system \textcolor{black}{and signal model} formulations for both the on-grid and off-grid scenarios. Section III \textcolor{black}{formalizes} the corresponding RDP recovery problems. Sections IV and V develop the proposed on-grid and off-grid recovery algorithms, respectively. Section VI reports numerical results, and evaluates the performance of the proposed methods. Finally, Section VII concludes the article.

\textit{Notation:} Vectors and matrices are denoted by boldface lower-case and upper-case letters, respectively\textcolor{black}{.} $\|\cdot \|$ denotes the Frobenius norm. $(\cdot )^T$, $(\cdot )^*$, $(\cdot )^H$, and $(\cdot )^{-1}$  are the transpose, conjugate, conjugate transpose and matrix inverse operators, respectively. The $\0_{m \times n}$ and $\I_n$ represent the $m \times n$ zero matrix and $n \times n$ identity matrix, respectively. $\mathbb{C}^{N}$ and $\mathbb{C}^{N,M}$ are respectively the sets of $N$-dimensional vectors of complex number, of $N \times M$ matrices of complex numbers. $\Re\{ \cdot\}$ and $\Im\{ \cdot\}$ mean the real and imaginary parts with $j=\sqrt{-1}$. $|\cdot|$ and $\angle\{\cdot\}$ are the magnitude and phase of a complex-valued scalar, respectively. \textcolor{black}{$\A(\mathcal{S}_r,:)$ and $\A(:,\mathcal{S}_c)$ denote the submatrices of $\A$ formed by selecting the rows and columns indexed by index sets $\mathcal{S}_r$ and $\mathcal{S}_c$, respectively.} $\A(n,:)$ and $\A(:,m)$ represent $n$-th row and $m$-th column of the matrix $\A$, respectively. $\a(n)$ is $n$-th element of the vector $\a$. $\delta(\cdot) $, $\otimes$ and $\odot$ denote the Kronecker delta function, Kronecker product and Hadamard product, respectively. $\text{vec}(\A)$ is a column vector consisting of all the columns of $\A$ stacked, \textcolor{black}{while $\text{unvec}(\a)$ represents its inverse operation}, which reshapes a vector $\a$ back into the corresponding matrix form. $\diag\{\a\}$ represents a diagonal matrix with vector $\a$ as its diagonal element. $\nabla_{a} f(a)|_{a=\bar{a}}$ denotes the gradient of the function $f(a)$ with respect to $a$, evaluated at $a=\bar{a}$. \textcolor{black}{$\text{Card}\{\mathcal{S}\}$ calculates the cardinality of the finite set $\mathcal{S}$.}

\section{System and signal model}

Consider a monostatic pulse-Doppler radar transmitting a coherent train of $N$ uniformly spaced pulses. After pulse compression, the radar collects data from $K_p$ consecutive range cells. The fast-time (range) index is denoted by $k=1,...,K_p$, and the $N$ slow-time measurements gathered at the $k$-th range cell are stacked in the vector $\z_k \in \mathbb{C}^{N}$. Let us assume that a swarm of scatterers, distributed over the range and Doppler domains, lies within the observed range swath. This swarm may represent either a distributed target or multiple discrete targets. Now denoting by  $D_k$ the number of scatterers belonging to the 
$k$-th range cell, the measurement vector $\z_k$ can be modeled as
\begin{align}
	\z_k = \sum_{d=1}^{D_k} \h(\tilde{f}_{d,k}) \tilde{x}_{d,k} + \n_{k}, \label{def_yl}
\end{align} 
where \newline
 $\bullet$ $\tilde{x}_{d,k}$ and $\tilde{f}_{d,k}$ indicate the \textcolor{black}{complex} echo strength and the Doppler frequency associated with the $d$-th \textcolor{black}{scatterer} of the $k$-th range cell.  \newline
$\bullet$ $\h(f)=[1,e^{j 2 \pi f T},...,e^{j 2 (N-1) \pi f T}]^T \in \mathbb{C}^{N}$ represents the Doppler steering vector associated with the Doppler frequency $f$, \textcolor{black}{and $T$ is the pulse repetition interval}; \newline
$\bullet$ $\{\n_{k}\}_{k=1}^{K_p}$ denote independent and identically distributed, complex, zero-mean, circular Gaussian random vectors \textcolor{black}{with covariance matrix $\M$} modeling the colored interference plus additive white Gaussian noise.

The received measurements can be conveniently represented in the range-Doppler domain, where each resolution  cell corresponds to the contribution of a potential scatterer characterized by a specific range delay and Doppler frequency. Owing to the limited number of dominant \textcolor{black}{scatterers}, the resulting  range-Doppler profile (RDP) is typically sparse, with only a few nonzero entries \cite{7484756,8708966}. To recover RDP, a uniform grid of $D$ Doppler frequencies $\{f_1,...,f_D\}$ spanning the unambiguous Doppler interval, is used, where $D$ is the size of the grid. \textcolor{black}{This grid is selected sufficiently dense such that there will be at most one scatterer within each bin $[f_d-\Delta f/2,f_d+\Delta f/2]$, where $\Delta f$ denotes the Doppler discretization step.}

\textcolor{black}{Accordingly, each target Doppler frequency is modeled as the closest grid point $f_{d}$ plus a small offset $\tilde{\epsilon}_{d,k}$, i.e.,
\begin{align}
\tilde{f}_{d,k} =f_{d}+ \tilde{\epsilon}_{d,k}, d=1,...,D;k=1,...,K_p, \label{def_fdk}
\end{align}}
where $|\tilde{\epsilon}_{d,k}| \leq \Delta f/2$.  \textcolor{black}{Depending on whether $\tilde{\epsilon}_{d,k}$ equals zero, the problem of RDP recovery can be categorized into two cases, namely the on grid and off grid setups detailed below.}
	
(i) When $\tilde{\epsilon}_{d,k}=0$, the scenario corresponds to the on-grid case, in which the Doppler frequency of each scatterer is a specific point of the predefined grid. Hence, the signal model in \eqref{def_yl}, can be expressed in the following form
\begin{align}
	\z_k = \sum_{d=1}^{D}  \h(f_d) {x}_{d,k} + \n_{k},k=1,...,K_p, \label{def_yl_overal}
\end{align}
where $\{\h(f_d)\}_{d=1}^{D}$, define the overall dictionary, and $\{{x}_{d,k}\}_{\{d=1,...,D;k=1,...,K_p\}}$ represent the unknown RDP in the on-grid case. Specifically, the value of ${x}_{d^{\star},k^{\star}}$ depends on whether a target occupies the range-Doppler index pair $(d^{\star},k^{\star})$. \textcolor{black}{If no target is present at $(d^{\star},k^{\star})$, the corresponding coefficient is assumed to be zero, i.e., ${x}_{d^{\star},k^{\star}}=0$.} Conversely, if a target is present at this index pair, the corresponding coefficient is nonzero and ${x}_{d^{\star},k^{\star}}$ is related to the scattering strength of the target, \textcolor{black}{as well to other terms involved in the radar range equation.}

Therefore, introducing the overall Doppler steering dictionary matrix $\H=[\h(f_1),\h(f_2),...,\h(f_D)]\in \mathbb{C}^{N,D}$, \eqref{def_yl_overal}  can be rewritten as
\begin{align}
\z_k =   \H \x_{k} + \n_{k},k=1,...,K_p, \label{def_yHxn_overal}
\end{align}
where $\x_{k}=\[{x}_{1,k},{x}_{2,k},...,{x}_{D,k}\]^T\in \mathbb{C}^{D}$, $k=1,...,K_p$. In a compact matrix form, \eqref{def_yHxn_overal} boils down to
\begin{align}
\Z =  \H \X +\N, \label{def_Z}
\end{align}
where 
\begin{align}
&\Z  = [\z_1,...,\z_{K_{p}}] \in \mathbb{C}^{N,K_p}, \\
&\X = [\x_1,...,\x_{K_{p}}] \in \mathbb{C}^{D,K_p}, \\
&\N = [\n_{1},...,\n_{K_p}] \in \mathbb{C}^{N,K_p}.
\end{align}

(ii) When $\tilde{\epsilon}_{d,k} \neq 0$, \textcolor{black}{the off-grid situation arises. This is tantamount to considering the presence of deviations/displacements between a scatterer Doppler frequency and the assumed grid.} The corresponding signal model can be expressed as
\begin{align}
	\bar{\z}_k = \sum_{d=1}^{D}  {\h}(f_d+\epsilon_{d,k}) {x}_{d,k} + \n_{k}, \label{signal_model_off}
\end{align}
where ${\h}(f_d+\epsilon_{d,k})$ and $\epsilon_{d,k}$, $d=1,...,D$, define the overall dictionary and the corresponding offsets at the $k$-th range index, respectively. 

It is assumed that the interval $[f_{d}-\Delta f/2,f_{d}+\Delta f/2]$ is sufficiently small such that the Doppler steering vector within this region lies in the validity \textcolor{black}{domain} of a first-order Taylor approximation around $f_{d}$ \cite{11204963,4385788}. Specifically,
	\begin{align}
		\h(\tilde{f}_{d,k}) \simeq \h(f_{d}) + \dot{\h}(f_{d}) \tilde{\epsilon}_{d,k}, \label{def_1st_tayler}
	\end{align}
	where $\dot{\h}(f_{d})$ denotes the derivative of $\h(f_{d})$ with respect to $f_d$ evaluated at  $f_{d}$ \cite{9272871}, given by
	\begin{align}
		\h(f_{d}) \odot j2\pi T[0,1,..,N-1]^T.
\end{align}

At the $k$-th fast-time index, the off-grid overall dictionary matrix $\H_{\bep_k} \in \mathbb{C}^{N,D}$ parameterized by the deviation vector $\bep_k =[\epsilon_{1,k},...,\epsilon_{D,k}]^T \in \mathbb{R}^{D} $ is denoted by
\textcolor{black}{
	\begin{align}
		&\H_{\bep_k} = [\h(\tilde{f}_{1,k}),\h(\tilde{f}_{2,k}),...,\h(\tilde{f}_{D,k})] \in \mathbb{C}^{N,D},  \label{def_H_bepk} \\ 
		&\simeq \bar{\H}_{\bep_k} = [\h(f_1) + \dot{\h}(f_1) \epsilon_{1,k},\h(f_2) + \dot{\h}(f_2) \epsilon_{2,k},..., \nonumber \\
		&~~~~~~~~~~~~\h(f_D) + \dot{\h}(f_D) \epsilon_{D,k}], \nonumber \\
		& = \H + \dot{\H} \diag\{\bep_k\}, \label{def_barH_HHeps}
\end{align}}
where $f_d=(d-1)\Delta f$, $\H = [\h(f_1),\h(f_2),...,\h(f_D)]$ and $\dot{\H}=[\dot{\h}(f_1),\dot{\h}(f_2),...,\dot{\h}(f_D)]$

In the off-grid scenario, the overall dictionary depends on the fast time index $k$ through the frequency displacement vector $\bep_k$, in general leading to $K_p$ \textcolor{black}{distinct} dictionaries.
Therefore, the signal model in \eqref{signal_model_off} can be compactly rewritten as
\begin{align}
	\bar{\z}_k = \bar{\H}_{\bep_k} \x_k+\n_k,k=1,...,K_p. \label{def_barzk}
\end{align}

\section{Problem Formulation} 

In this section, leveraging on the developed signal models, the RDP recovery problems are addressed. Specifically, the RDP is assumed to be sparse, implying that most entries of $\X$ are zero. This sparsity structure will be exploited to develop a specific recovery framework.

\subsection{On-Grid Recovery}

To \textcolor{black}{retrieve} the RDP in the on-grid case from the received data, the statistical model of the observations is first capitalized by writing the probability density function (PDF) of the received data
\begin{align}
	&p_1(\Z;\X,\M) = \frac{1}{[\pi^{N}\det(\M)]^{K_p}}\times  \nonumber \\
	&\exp\( -\sum_{k=1}^{K_p} (\z_k-\H\x_k)^H \M^{-1} (\z_k-\H\x_k)  \). \label{p1}
\end{align}

\textcolor{black}{In the following, it} is assumed that the radar system operates in a standard homogeneous environment, such that a secondary data matrix $\Z_s \in \mathbb{C}^{N,K_s},K_s \geq N$ can be collected in target-free range cells \cite{richmond2002performance,10077421}. Specifically, the $K_s$ columns of $\Z_s$ are modeled as independent and identically distributed complex, zero-mean, circularly symmetric Gaussian random vectors with covariance matrix $\M$ \cite{9272871}. Accordingly, 
\begin{align}
	p_0(\Z_s;\M) = \frac{1}{[\pi^{N}\det(\M)]^{K_s}} \exp\( -\tr( \Z_s^H \M^{-1} \Z_s) \). \label{p0}
\end{align}

The joint PDF of the primary and secondary data (under the statistical independence assumption between $\Z$ and $\Z_s$) is given by
\begin{align}
	&p(\Z,\Z_s;\X,\M) = \frac{1}{(\pi^N \det(\M))^{(K_p+K_s)}} \times \nonumber \\
	& \exp\( 
	- \sum_{k=1}^{K_p} (\z_k-\H\x_k)^H \M^{-1} (\z_k-\H\x_k)
	- \tr(\Z_s^H \M^{-1}\Z_s )
	\).
\end{align}

To perform RDP inference while accounting for scatterer sparsity and the presence of the nuisance parameter $\M$, either the one-step or the two-step regularized maximum likelihood (RML) estimation approach can be adopted \cite{7126163,9699069}, as carefully illustrated below. 
Following the RML estimation framework in \cite{5424069,8574991}, an element-sparsity-promoting penalty function is introduced firstly
\textcolor{black}{
\begin{align}
	f_1(\X) =  \sum_{d=1}^{D} \sum_{k=1}^{K_p} \frac{2}{q} \[ (| x_{d,k} |^2+\gamma)^{\frac{q}{2}}-1 \], \label{def_f1_X} 
\end{align}
where $\gamma>0$ is the smoothing factor making \eqref{def_f1_X} differentiable, whereas the parameter $0<q\leq 1$ rules the $l_q$-norm–based sparsity penalty applied to the range–Doppler cell energy.}

Two strategies, namely the one-step and the two-step approaches, can be employed for RDP recovery \cite{7126163,9699069}. The former jointly estimates the interference-plus-noise covariance matrix $\M$ and the RDP $\X$.
The latter first replaces the unknown covariance matrix $\M$ with the sample covariance matrix (SCM) $\hat{\M}$, computed from the target-free secondary data $\Z_s$ \cite{4655460,7345553,10508618,8052571,6166345}, then the RDP recovery is performed. The details are given below.

\subsubsection{One-step recovery} 
The following regularized minimization problem with respect to $\M$ and $\X$ for one-step sparse RDP recovery is formulated \textcolor{black}{as}
\begin{align}
	&\min_{\X,\M}~(K_p+K_s)\log(\pi^N \det(\M)) \nonumber \\
	&+ \sum_{k=1}^{K_p} (\z_k-\H\x_k)^H \M^{-1} (\z_k-\H\x_k) \nonumber \\
	&+ \tr(\Z_s^H \M^{-1}\Z_s )+f_1(\X), \label{def_max_e}
\end{align}
and the corresponding algorithm to tackle \eqref{def_max_e} is referred \textcolor{black}{to} the one-step RDP on-grid recovery (ORDP-on) algorithm.

\subsubsection{Two-step recovery} 
Here, the SCM is computed first, i.e., $\hat{\M}=\frac{1}{K_s}\Z_s\Z_s^H$. \textcolor{black}{Then}, ignoring the constant terms functionally independent of $\X$, the regularized two-step sparse RDP recovery is formulated as
	\begin{align}
		&\min_{\X}~   \sum_{k=1}^{K_p} (\z_k-\H\x_k)^H \hat{\M}^{-1} (\z_k-\H\x_k)+f_1(\X). \label{def_max_e_scm}
	\end{align}

The algorithm to handle \eqref{def_max_e_scm} is referred to as the two-step RDP on-grid recovery algorithm (TRDP-on).

\subsection{Off-Grid Recovery}

Still assuming the presence of secondary data $\Z_s$, the joint PDF \textcolor{black}{$p(\bar{\Z},\Z_s;\X,\M,\{ \bep_k\}_{k=1}^{K_p})$ under the off-grid assumption can be written as
	\begin{align}
		&p(\bar{\Z},\Z_s;\X,\M,\{ \bep_k\}_{k=1}^{K_p}) = \frac{1}{(\pi^N \det(\M))^{(K_p+K_s)}} \nonumber \\
		&\times \exp(  
		-\sum_{k=1}^{K_p} (\bar{\z}_k-\bar{\H}_{\bep_k}\x_k)^H \M^{-1} (\bar{\z}_k-\bar{\H}_{\bep_k}\x_k)\nonumber \\
		&~~~~~~~~~~~~~~~~ - \tr(\Z_s^H \M^{-1}\Z_s )	). 
\end{align}}

Like in the on-grid setup, the range-Doppler profile recovery process can be formulated according to either the one-step or the two-step \textcolor{black}{approaches}, as detailed below.
\subsubsection{One-step recovery}

\textcolor{black}{The problem of jointly estimating $\{\M,\X,\{\bep_k\}_{k=1}^{K_p}\}$ can be cast as}
\textcolor{black}{
\begin{align}
	&\min_{\M,\X,\{\bep_k\}_{k=1}^{K_p}} ~~ (K_p+K_s)\log(\pi^N \det(\M)) \nonumber \\ 
	&+ \sum_{k=1}^{K_p} (\bar{\z}_k-\bar{\H}_{\bep_k}\x_k)^H \M^{-1} (\bar{\z}_k-\bar{\H}_{\bep_k}\x_k) \nonumber \\
	&+ \tr(\Z_s^H \M^{-1}\Z_s )+f_1(\X) \nonumber  \\
	&\text{s.t.}~~|\epsilon_{d,k}| \leq \Delta f/2,   d = 1,...,D;k=1,...,K_p.\label{def_max_offG}
\end{align}}
\textcolor{black}{The resulting algorithm \textcolor{black}{to solve \eqref{def_max_offG} is denoted} as the one-step RDP off-grid recovery algorithm (ORDP-off).}

\subsubsection{Two-step recovery}\textcolor{black}{The joint retrieval of $\X$ and $\{\bep_k\}_{k=1}^{K_p}$ is formulated as}
\begin{align}
	&\min_{\X,\{\bep_k\}_{k=1}^{K_p}}~~\sum_{k=1}^{K_p} (\bar{\z}_k-\bar{\H}_{\bep_k}\x_k)^H \hat{\M}^{-1} (\bar{\z}_k-\bar{\H}_{\bep_k}\x_k)
	\nonumber \\
	&+f_1(\X) \nonumber \\
	&\text{s.t.}~~~~|\epsilon_{d,k}| \leq \Delta f/2,   d = 1,...,D;k=1,...,K_p.	 \label{def_max_offGSCM}
\end{align}
The algorithm to solve \eqref{def_max_offGSCM} is referred to the two-step RDP off-grid recovery algorithm (TRDP-off).

\section{On-grid Case RDP Recovery Procedures}
\label{on_grid_alg}

This section focuses on solving \textcolor{black}{Problem} \eqref{def_max_e} first. Then, the solution to \eqref{def_max_e_scm} is provided.

\subsection{ORDP-on algorithm}

Problem \eqref{def_max_e} can be equivalently rewritten as
\textcolor{black}{
\begin{align}
	&\min_{\X,\M}~(K_p+K_s) \log(\det(\M)) \nonumber \\
	&+ \tr \(	\M^{-1}\sum_{k=1}^{K_p}   (\z_k-\H\x_k)(\z_k-\H\x_k)^H
	 +  \M^{-1} \Z_s  \Z_s^H  \) \nonumber \\
	&+f_1(\X). \label{def_max_log}
\end{align}}

Letting
\begin{align}
	\S(\X)=\sum_{k=1}^{K_p}   (\z_k-\H\x_k)(\z_k-\H\x_k)^H 
	 + \Z_s \Z_s^H, \label{def_SX}
\end{align}
the optimization problem in \eqref{def_max_log} involving both $\M$ and $\X$, can be rewritten as \cite{5424069,8574991}
\begin{align}
	&\min_{\X,\M} (K_p+K_s) \log(\det(\M)) + \tr(\M^{-1} \S(\X)  )  +f_1(\X). \label{def_max_log_f1}
\end{align}

\textcolor{black}{Optimizing over $\M$ yields}
\begin{align}
	{\M}=\frac{\S(\X)}{K_p+K_s}. \label{solutiontoW}
\end{align}

Substituting \eqref{solutiontoW} into \eqref{def_max_log_f1} leads to the concentrated optimization problem
\begin{align}
	&\min_{\X} (K_p+K_s) \log(\det(\S(\X)))+f_1(\X). \label{pro_X_only}
\end{align}

To \textcolor{black}{deal with} \eqref{pro_X_only}, the majorization-minimization (MM) approach is adopted, and the objective function is defined as $h(\X)$. \textcolor{black}{Precisely}, the majorized function of $\log(\det(\S(\X)))$ is developed firstly \cite{10715588,hunter2004tutorial,7547360}, i.e.,
\begin{align}
	&\log(\det(\S(\X))) \leq \log(\det(\S(\bar{\X}))) \nonumber \\
	&+\tr \[ (\nabla_{\Y} \{\log(\det(\Y))\}|_{\Y=\S(\bar{\X})})^H( \S(\X)-\S(\bar{\X}) ) \], \label{def_logdetSX}
\end{align}
where $\bar{\X}$ is a given point in the feasible set (e.g., the solution obtained at the previous iteration) \cite{8574991}.

\textcolor{black}{Due to 
\begin{align}
	\nabla_{\Y} \{\log(\det( \Y ))\}|_{\Y=\S(\bar{\X})} = \(\S({\bar{\X}})^{-1}\)^H= \S({\bar{\X}})^{-1},
\end{align}
\eqref{def_logdetSX} can be rewritten as
\begin{align}
	&\log(\det(\S(\X)))  \leq  \log(\det(\S(\bar{\X}))) +\tr(\S(\bar{\X})^{-1} \S(\X))-N, \label{MM_lambda}
\end{align}
\textcolor{black}{with equality achieved at} $\bar{\X}=\X$.}

The majorizer of the element-sparsity promoting penalty $f_1(\X)$ is now developed. \textcolor{black}{Since this regularization term} is composed of $D \times K_p$ elements, the corresponding majorizing function can be obtained via a suitable majorization of each term $(| x_{d,k} |^2+\gamma)^{\frac{q}{2}}$. According to \cite{hunter2004tutorial,7547360} 
\begin{align}
	&( | x_{d,k} |^2 +\gamma)^{\frac{q}{2}} \leq (  | \bar{x}_{d,k} |^2+\gamma)^{\frac{q}{2}} \nonumber \\
	&+\frac{q}{2}(| \bar{x}_{d,k} |^2+\gamma)^{\frac{q}{2}-1}(| {x}_{d,k} |^2-| \bar{x}_{d,k} |^2),
\end{align}
\textcolor{black}{with equality achieved at $\bar{x}_{d,k}=x_{d,k}$. Thus, the majorizing function of the penalty term $\frac{2}{q} [ (| x_{d,k} |^2+\gamma)^{\frac{q}{2}}-1 ]$ is given by}
\begin{align}
	&\frac{2}{q}\[(| \bar{x}_{d,k}|^2+\gamma)^{\frac{q}{2}}-1\]+\bar{c}_{d,k}| x_{d,k} |^2-\bar{c}_{d,k}| \bar{x}_{d,k} |^2, \label{major_X}
\end{align}
where 
\begin{align}
		\bar{c}_{d,k} = (| \bar{x}_{d,k} |^2+\gamma)^{\frac{q}{2}-1}. \label{def_cdk}
\end{align}

As a consequence, \textcolor{black}{a surrogate function of $f_1(\X)$ is}
\begin{align}
	&f_1(\X) \leq \sum_{d=1}^{D} \sum_{k=1}^{K_p} \left\{ \frac{2}{q}\[(| \bar{x}_{d,k}|^2+\gamma)^{\frac{q}{2}}-1\] \right. \nonumber\\
	&\left.+\bar{c}_{d,k}| x_{d,k} |^2-\bar{c}_{d,k}| \bar{x}_{d,k} |^2\right\}. \label{major_f1}
\end{align}

Substituting \eqref{MM_lambda} and \eqref{major_f1} into \eqref{pro_X_only}, and discarding the terms that are functionally independent of $\X$, yields the following surrogate minimization problem
\begin{align}
	\min_{\X}~(K_p+K_s)\tr(\S(\bar{\X})^{-1} \S(\X)) + \sum_{d=1}^{D} \sum_{k=1}^{K_p}  \bar{c}_{d,k}| x_{d,k} |^2. \label{def_lam_cdk_x}
\end{align}

The following proposition summarizes the \textcolor{black}{the key theoretical features of the proposed solution technique.}

\textcolor{black}{\textbf{Proposition 1:}} Let $\X^{(0)}$ be a feasible solution to \textcolor{black}{Problem} \eqref{pro_X_only}. Moreover, let $\X^{(e+1)}$ be the optimal solution to \textcolor{black}{Problem} \eqref{def_lam_cdk_x} \textcolor{black}{when $\bar{\X}=\X^{(e)}$, with $e \in \mathbb{N}$ representing} the iteration number (IN). Then \newline
$\bullet$ \textcolor{black}{the sequence of objective function values in} \eqref{pro_X_only} induced by the generated sequence of points $\X^{(e)}$ monotonically decreases; \newline
$\bullet$ any cluster point of the produced bounded sequence $\X^{(e)}$ is a KKT point to \eqref{pro_X_only}. \newline
\textit{Proof:} See Appendices A and B in the supplementary material.

To facilitate the subsequent derivation, $\S(\bar{\X})$ is factorized as
\begin{align}
	\S(\bar{\X})^{-1} = \tilde{\M}^H \tilde{\M},
\end{align}
i.e., $\tilde{\M}=\S(\bar{\X})^{-\frac{1}{2}}$; thus $\tr(\S(\bar{\X})^{-1} \S(\X))$ can be cast as
\begin{align}
	&\sum_{k=1}^{K_p} \| \tilde{\z}_k-\tilde{\H} \x_k\|^2 + \| \tilde{\M} \Z_s \|^2,
\end{align}
where 
\begin{align}
	&\tilde{\z}_k = \tilde{\M}\z_k, \label{def_tZ} \\
	&\tilde{\H} = \tilde{\M}\H. \label{def_tH}
\end{align}

\textcolor{black}{
Hence, introducing
\begin{align}
	&\x = \text{vec}(\X), \\
	&\tilde{\z} = \text{vec}(\tilde{\Z})=\text{vec}([\tilde{\z}_1,\ldots,\tilde{\z}_{K_p}]), \\
	&\C = \diag\{ \bar{c}_{1,1},\bar{c}_{2,1},...,\bar{c}_{D,K_p} \}.
\end{align}}

\eqref{def_lam_cdk_x} is reformulated as
\begin{align}
	&\min_{\x}~	(K_p+K_s) \| \tilde{\z}-(\I_{K_p} \otimes \tilde{\H}) \x \|^2 +  \x^H \C \x. \label{zHx}
\end{align}

\textcolor{black}{whose optimizer is}
\begin{align}
	\x	=\((\I_{K_p} \otimes \tilde{\H}^H\tilde{\H} )  +  \frac{\C}{(K_p+K_s)} \)^{-1}(\I_{K_p} \otimes \tilde{\H})^H\tilde{\z} . \label{def_x}
\end{align}

The developed ORDP-on \textcolor{black}{technique} is summarized in \textbf{Algorithm} \ref{RDP-on}, which is executed until convergence (or a maximum IN) is reached. \textcolor{black}{A matched filter initialization is considered}, i.e., $\X^{(0)}= \diag(\| \H(:,1)\|,...,\| \H(:,D)\|)^{-2} \H^H\Z$, which projects the observations onto the column space of $\mathbf{H}$.

\subsection{TRDP-on algorithm}


\textcolor{black}{
According to the discussion leading to \eqref{def_max_log_f1}-\eqref{major_X}, a surrogate function of \eqref{def_max_e_scm} is constructed as
\begin{align}
	&\min_{\x}~\sum_{k=1}^{K_p} (\z_k-\H\x_k)^H \hat{\M}^{-1} (\z_k-\H\x_k) + \sum_{d=1}^D \sum_{k=1}^{K_p} \bar{c}_{d,k}| x_{d,k} |^2,\label{min_XZHX_S}
\end{align}}
where the definition of $ \bar{c}_{d,k}$ is given by \eqref{def_cdk}.

Problem \eqref{min_XZHX_S} can be cast as
\textcolor{black}{
\begin{align}
	\min_{\x}~\sum_{k=1}^{K_p} \|\dot{\z}_k-\check{\H}\x_k\|^2
	+ \sum_{d=1}^D \sum_{k=1}^{K_p} \bar{c}_{d,k}| x_{d,k} |^2, \label{ZHX_M}
\end{align}}
\textcolor{black}{where
\begin{align}
	&\hat{\M}^{-1}=\[\frac{1}{K_s} \Z_s\Z_s^H\]^{-1} = \check{\M}^H \check{\M}, 
\end{align} 
and
 \vspace{-0.3cm}
\begin{align}
	&\dot{\z}_k= \check{\M}\z_k ,\\
	&\check{\H} = \check{\M}\H.
\end{align}}
 \vspace{-0.5cm}

The solution to \eqref{ZHX_M} is given by
\textcolor{black}{
\begin{align}
	\x	=\((\I_{K_p} \otimes \check{\H}^H\check{\H} )  +  \C \)^{-1}(\I_{K_p} \otimes \check{\H})^H\dot{\z},
\end{align}}
where $\dot{\z} = [\dot{\z}_1^T,\dot{\z}_2^T,...,\dot{\z}_{K_p}^T]^T$.

Compared with the ORDP-on, the TRDP-on algorithm is simpler since it relies on a given SCM. Consequently, its implementation procedure and convergence conditions are consistent with those of Algorithm 1, and are not repeated here for brevity.

\begin{algorithm}[h]
	\caption{ ORDP-on algorithm} \label{RDP-on}
	\KwIn{$\H$, $\Z$, $\Z_s$, $D$, $K_p$, $\gamma$, $q \in (0,1)$, maximum IN $e_{\text{max}}$, convergence tolerance $\xi$}.
	\textbf{Initialization:} Set $e=0$ and $\X^{(0)}= \diag(\| \H(:,1)\|,...,\| \H(:,D)\|)^{-2} \H^H\Z$. \\
	\Repeat{$|h(\X^{(e)}) -h(\X^{(e-1)})|  \leq \xi$ or $e > e_{\text{max}}$}{
		Assign $\bar{\X}=\X^{(e)}$. \\
		Compute $\S(\bar{\X})=\sum_{k=1}^{K_p}   (\z_k-\H\bar{\x}_k)(\z_k-\H\bar{\x}_k)^H 
		+ \Z_s \Z_s^H$.\\
		Compute $\bar{c}_{d,k} = (| \bar{x}_{d,k} |^2+\gamma)^{\frac{q}{2}-1},d=1,...,D; k=1,...,K_p$. $\C = \diag\{ \bar{c}_{1,1},\bar{c}_{2,1},...,\bar{c}_{D,K_p} \}$. \\
		Determine $\tilde{\Z}$ and $\tilde{\H}$ via \eqref{def_tZ} and \eqref{def_tH}, respectively. \\
		Compute $\x	=\((\I_{K_p} \otimes \tilde{\H}^H\tilde{\H})  +  \frac{\C}{(K_p+K_s)}\)^{-1}(\I_{K_p} \otimes \tilde{\H})^H\tilde{\z} $.  $\X^{(e+1)}=\textcolor{black}{\text{unvec}(\x)}$. \\
		Set $e=e+1$.}
	\KwOut{\textcolor{black}{Estimated RDP $\hat{\X}$ = $\X^{(e)}$.} }
\end{algorithm}

\subsection{On-grid Case RDP Refinement Strategy}

Leveraging the estimate $\hat{\X}$ obtained from the ORDP-on or TRDP-on algorithm, \textcolor{black}{an effective procedure for refining the RDP is summarized as follows.}

\textbf{Step 1:} Identification of the active atoms in the overall Doppler steering dictionary matrix $\H$ via a model selection criterion, \textcolor{black}{yielding an estimate $\mathcal{I}$} of the target support.

\textbf{Step 2:} Estimation of the refined RDP $\breve{\X}$ according to the maximum-likelihood (ML) criterion, \textcolor{black}{with the active bins/atoms obtained in Step 1.}

Step 1 relies on the BIC framework \cite{8574991,stoica2004model,roberts2010iterative}. Let $\grave{\x} \in \mathbb{C}^{D K_p}$ be the vector obtained by vectorizing $\hat{\X}$ (e.g., via column-wise stacking) and sorting its entries in descending order of magnitude, i.e.,
\begin{align}
	| \grave{\x}(1) |^2 \geq | \grave{\x}(2) |^2 \geq \cdots \geq | \grave{\x}(DK_p) |^2. \label{def_gravX}
\end{align}

Let $\{(r_i, c_i)\}_{i=1}^{DK_p}$ denote the corresponding row and column indices of the entries of $\hat{\X}$ associated with $\grave{\x}(i)$, i.e.,
\begin{align}
	\grave{\x}(i) = \hat{\X}(r_i, c_i), \quad i=1,\ldots,DK_p. \label{def_u_set}
\end{align}

Then $\mathcal{I}$ is defined as
\begin{align}
	\mathcal{I}=\{1,...,i^{\star}\}, \label{def_R} 
\end{align}
where  
\begin{align}
	i^{\star}=\arg \min_{i \in \{ 1,...,\bar{i} \}} \text{BIC}(i), \label{def_i_star}
\end{align}
with 
	\begin{align}
		&\text{BIC}(i) = \nonumber \\
		&2 (K_p+K_s) \log\left( \det\left(  \sum_{k =1}^{K_p}(\z_k   - \dot{\H}_k^{i} \dot{\x}_k^{i})(\z_k   - \dot{\H}_k^{i} \dot{\x}_k^{i})^H \right. \right. \nonumber \\
		&\left.\left. +\Z_s \Z_s^H  \right) \right) +(4i +N^2) \log(2 N (K_p+K_s)). \label{def_BIC_i_on}
\end{align} 
Moreover, for any given $k$ and $i$, let 
\begin{align}
	\mathcal{S}_k(i) = \{ \tilde{r} \mid \exists~ l \in \{1,...,i\}, \text{such that}~(\tilde{r},k)=(r_l,c_l) \} \label{def_S_k}
\end{align}
denote the set of row indices of the first $i$ selected pairs $\{(r_l, c_l)\}_{l=1}^{i}$ whose column index equals $k$, and \newline
$\bullet$ $\dot{\H}_k^{i}$ and $\dot{\x}_k^{i}$ are defined as
\begin{align}
	&\dot{\H}_k^{i} = \H(:,\mathcal{S}_k(i)) \in \mathbb{C}^{N, \text{Card}\{\mathcal{S}_k(i)\}}, \\
	&\dot{\x}_k^{i} = \hat{\X}(\mathcal{S}_k(i),k) \in \mathbb{C}^{ \text{Card}\{\mathcal{S}_k(i)\}},
\end{align}
\textcolor{black}{and when $\mathcal{S}_k(i) =\emptyset $, $\dot{\H}_k^{i} \dot{\x}_k^{i}=\0$;}\newline
$\bullet$\textcolor{black}{$\bar{i}$ is a given upper bound to the actual scatterer number, determined based on prior information or system design considerations ($\bar{i}=\frac{K_p}{2}$ for this work)}; \newline
$\bullet$ the second term of \eqref{def_BIC_i_on} represents the BIC-based penalty \cite{8574991,stoica2004model,roberts2010iterative}, $NK_p$ is the size of the observation data,  $4i+N^2$ accounts for the number of real-valued unknowns when $i$ is the assumed number of unknown scatterers: amplitude (complex valued), range and Doppler \cite{roberts2010iterative}, and $N^2$ unknown real parameters in unknown covariance matrix estimation.

For the on-grid case, the RDP refinement procedure shares the same framework as the off-grid formulation presented in Appendix C of the supplemental material. Specifically, the former corresponds to the special case of the latter in which off-grid frequency displacements vector $\bep$ is fixed to zero. Therefore, the on-grid RDP refinement (accomplished by this Step 2) can be performed by applying Step 3 of the off-grid procedure with $\bep=0$, and thus it is omitted here for brevity.

\section{Off-grid Case RDP Recovery Procedures}
\label{off_grid_alg}

\textcolor{black}{In this section, the solution to \eqref{def_max_offG} is presented. That for \eqref{def_max_offGSCM} is based on the same guidelines.}

\subsection{ORDP-off algorithm} 


Like in \eqref{def_SX}, the function of $\X$ and $\{\bep_k\}_{k=1}^{K_p}$ is defined firstly
\begin{align}
	\bar{\S}(\X,\{\bep_k\}_{k=1}^{K_p}) = \sum_{k=1}^{K_p}(\bar{\z}_k-\bar{\H}_{\bep_k}\x_k)(\bar{\z}_k-\bar{\H}_{\bep_k}\x_k)^H+\Z_s  \Z_s^H, \label{def_barS}
\end{align}

Hence, \textcolor{black}{Problem} \eqref{def_max_offG} can be cast into
\begin{align}
	&\min_{\X,\{\bep_k\}_{k=1}^{K_p}} (K_p+K_s) \log(\det(\bar{\S}(\X,\{\bep_k\}_{k=1}^{K_p})))+f_1(\X) \nonumber \\
	&~~\text{s.t.}~~|\epsilon_{d,k}| \leq \Delta f/2,   d = 1,...,D; k=1,...,K_p. \label{pro_barX_bep}
\end{align}

\textcolor{black}{Like in \eqref{def_logdetSX}, a surrogate function of $\log(\det(\bar{\S}(\X,\{\bep_k\}_{k=1}^{K_p})))$ is given by \cite{hunter2004tutorial,7547360}}
\begin{align}
	&\log(\det(\bar{\S}(\X,\{\bep_k\}_{k=1}^{K_p} ))) \leq \log(\det(\bar{\S}(\bar{\X},\{\bar{\bep}_k\}_{k=1}^{K_p}))) \nonumber \\
	&+ \tr[(\nabla_{\Y} \{\log(\det(\Y))\} |_{\Y=\bar{\S}(\bar{\X},\{\bar{\bep}_k\}_{k=1}^{K_p} )})^H  \nonumber \\
	&(\bar{\S}(\X,\{\bep_k\}_{k=1}^{K_p})-\bar{\S}(\bar{\X},\{\bar{\bep}_k\}_{k=1}^{K_p}))], \label{def_logdetbarX_bep}
\end{align}
where $\bar{\X}$ and $\{\bar{\bep}_k\}_{k=1}^{K_p}$ are given points in the feasible set (e.g., the solution obtained at the previous iteration) \cite{8574991}, and 
\begin{align}
	&\nabla_{\Y} \{\log(\det(\Y))\} |_{\Y=\bar{\S}(\bar{\X},\{\bar{\bep}_k\}_{k=1}^{K_p})}\nonumber \\
	&= (\bar{\S}(\bar{\X},\{\bar{\bep}_k\}_{k=1}^{K_p})^{-1})^H= \bar{\S}(\bar{\X},\{\bar{\bep}_k\}_{k=1}^{K_p})^{-1},
\end{align}
thus \eqref{def_logdetbarX_bep} can be cast as
\begin{align}
	&\log(\det(\bar{\S}(\X,\{\bep_k\}_{k=1}^{K_p}))) \leq \log(\det(\bar{\S}(\bar{\X},\{\bar{\bep}_k\}_{k=1}^{K_p}))) \nonumber \\
	&+\tr[ \bar{\S}(\bar{\X},\{\bar{\bep}_k\}_{k=1}^{K_p})^{-1} (\bar{\S}(\X,\{\bep_k\}_{k=1}^{K_p})-\bar{\S}(\bar{\X},\{\bar{\bep}_k\}_{k=1}^{K_p})) ]. \label{def_X_bepMM}
\end{align}
\textcolor{black}{with equality achieved at $\bar{\X}=\X$ and $\bar{\bep}_k=\bep_k,k=1,...,K_p$.}

Ignoring the terms that are functionally independent of $\X$ and $\{ \bep_k \}_{k=1}^{K_p}$, \eqref{pro_barX_bep} is \textcolor{black}{cast} as
\begin{align}
	&\min_{\X,\{ \bep_k \}_{k=1}^{K_p} }~(K_p+K_s)\tr(\bar{\S}(\bar{\X},\{\bar{\bep}_k\}_{k=1}^{K_p})^{-1} \bar{\S}(\X,\{\bep_k\}_{k=1}^{K_p} )) \nonumber \\ 
	&~~~~~~~~+ \sum_{d=1}^{D} \sum_{k=1}^{K_p}  \bar{c}_{d,k}| x_{d,k} |^2, \nonumber \\
	&~~\text{s.t.}~~|\epsilon_{d,k}| \leq \Delta f/2,   d = 1,...,D; k=1,...,K_p.\label{def_barSSX}
\end{align}

Here, introducing $\bar{\S}(\bar{\X},\{\bar{\bep}_k\}_{k=1}^{K_p})^{-1} = \bar{\M}^H \bar{\M} $ and ignoring constant terms, \eqref{def_barSSX} is rewritten as follows
\begin{align}
&\min_{\X,\{ \bep_k \}_{k=1}^{K_p}}~(K_p+K_s) \sum_{k=1}^{K_p}\|\bar{\M}\bar{\z}_k-\bar{\M}\bar{\H}_{\bep_k}\x_k\|^2 \nonumber \\
&+ \sum_{d=1}^D \sum_{k=1}^{K_p} \bar{c}_{d,k}| x_{d,k} |^2,\nonumber \\
&~~\text{s.t.}~~|\epsilon_{d,k}| \leq \Delta f/2,   d = 1,...,D; k=1,...,K_p. \label{pro_MZHeps_X}
\end{align}

Since \eqref{pro_MZHeps_X} couples $\X$ and $\{ \bep_k \}_{k=1}^{K_p}$ nonlinearly, it can be solved resorting to the \textcolor{black}{MBI (minimum block improvement) paradigm \cite{11204963}}, where the RDP $\X$ and the off-grid frequency displacements $\{ \bep_k \}_{k=1}^{K_p}$ are considered as \textcolor{black}{two} variable blocks. 
\textcolor{black}{According to this optimization framework, the optimization variables are first divided into disjoint blocks. Then, the objective function (or its surrogate) is minimized with respect to one block while keeping unaltered the others. At each iteration, only the block ensuring the best result in terms of objective function is updated for the next step.} Therefore, the MBI scheme performs a \textcolor{black}{dynamic} update by choosing the optimized block that provides the greatest improvement at each iteration.

Specifically, define the variable set $\mathcal{Y}=\{\X,\{\bep_k\}_{k=1}^{K_p}\}$, and let $\mathcal{Y}^{(e)}=\{\X^{(e)},\{\bep_k^{(e)}\}_{k=1}^{K_p}\}$ denote its instance at the $e$-th iteration, with attained objective value $u\(\mathcal{Y}^{(e)}\)=(K_p+K_s) \sum_{k=1}^{K_p}\|\bar{\M}\bar{\z}_k-\bar{\M}\bar{\H}_{\bep_k^{(e)}}\x_k^{(e)}\|^2 + \sum_{d=1}^D \sum_{k=1}^{K_p} \bar{c}_{d,k}| x_{d,k}^{(e)} |^2$. At each iteration, two candidate updates are constructed by optimizing one variable block while keeping unaltered the others. Then, the MBI procedure just updates the block ensuring the best objective improvement.

\textcolor{black}{Then, the MBI procedure requires solving (possibly resorting to bespoke surrogate functions) the following two subproblems at each iteration to update the optimized solution \cite{10891687}, that is
\begin{align}
	\mathcal{P}_{\X^{e+1}} \left\{
	\begin{aligned}
		&\min_{\X}~(K_p+K_s) \sum_{k=1}^{K_p}\|\bar{\M}\bar{\z}_k-\bar{\M}\bar{\H}_{\bep_k^{(e)}}\x_k\|^2  \\
		&~~~~~~+ \sum_{d=1}^D \sum_{k=1}^{K_p} \bar{c}_{d,k}| x_{d,k} |^2, 
	\end{aligned}
	\right.
	\label{pro_MZH_X}
\end{align}
and 
\begin{align}
	\mathcal{P}_{\{\bep_{k}^{e+1}\}_{k=1}^{K_p}} \left\{
	\begin{aligned}
		&\min_{\{ \bep_k \}_{k=1}^{K_p}}~\sum_{k=1}^{K_p}\|\bar{\M}\bar{\z}_k-\bar{\M}\bar{\H}_{\bep_k}\x_k^{(e)}\|^2, \\
		&~~\text{s.t.}~~|\epsilon_{d,k}| \leq \Delta f/2,   d = 1,...,D; k=1,...,K_p,
	\end{aligned}
	\right.
	\label{pro_MZHeps}
\end{align}
the solutions to Problems $\mathcal{P}_{\X^{e+1}}$ and $\mathcal{P}_{\{\bep_{k }^{e+1}\}_{k=1}^{K_p}}$ are discussed below.}

(1) \textcolor{black}{Solution to Problem $\mathcal{P}_{\X^{e+1}}$: Following \eqref{zHx}-\eqref{def_x}, \eqref{pro_MZH_X} can be transformed as follows}
\begin{align}
	&\min_{\x}~	(K_p+K_s) \| \tilde{\z}- \tilde{\H}^{(e)}_{\epsilon} \x \|^2 +  \x^H \C_x \x. \label{zHx-off}
\end{align}
where 
\begin{align}
	&\tilde{\z} = [(\bar{\M}\bar{\z}_1)^T,...,(\bar{\M}\bar{\z}_{K_p})^T]^T, \label{def_tz_off} \\
	&\tilde{\H}^{(e)}_{\epsilon} = \begin{bmatrix}
		\bar{\M}\bar{\H}_{\bep_1^{(e)}} & & & \\
		&  \bar{\M}\bar{\H}_{\bep_2^{(e)}} & & \\
		& & \ddots&  \\
		& & & \bar{\M}\bar{\H}_{\bep_{K_p}^{(e)}}
	\end{bmatrix}, \label{def_tH_off}\\
	&\C_x = \diag\{ \bar{c}_{1,1},\bar{c}_{2,1},...,\bar{c}_{D,K_p} \}.
\end{align}

The optimal solution to \eqref{zHx-off} is given by
\begin{align}
	&\x^{(e+1)}	=\(\(\tilde{\H}_{\epsilon}^{(e)}\)^H \tilde{\H}^{(e)}_{\epsilon}  +  \frac{\C_x}{(K_p+K_s)}\)^{-1}  \(\tilde{\H}_{\epsilon}^{(e)}\)^H\tilde{\z} . \label{def_x_OFF}
\end{align}
achieving as objective $\hat{u}_1^{(e+1)}=(K_p+K_s) \sum_{k=1}^{K_p}\|\bar{\M}\bar{\z}_k-\bar{\M}\bar{\H}_{\bep_k^{(e)}}\x_k^{(e+1)}\|^2 + \sum_{d=1}^D \sum_{k=1}^{K_p} \bar{c}_{d,k}| x_{d,k}^{(e+1)} |^2$.

(2) \textcolor{black}{Solution to Problem $\mathcal{P}_{\{\bep_{k}^{e+1}\}_{k=1}^{K_p}}$: the objective function in \eqref{pro_MZHeps} can be cast as}
\begin{align}
	&\|\tilde{\z}_x^{(e)} - \tilde{\H}_x^{(e)} \bep \|^2,
\end{align}
where 
\textcolor{black}{
\begin{align}
	&\tilde{\z}_x^{(e)} = \[ (\bar{\M}\bar{\z}_1-\bar{\M}\H\x_1^{(e)})^T,...,(\bar{\M}\bar{\z}_{K_p}- \bar{\M}\H\x_{K_p}^{(e)})^T \]^T, \\
	&\tilde{\H}_x^{(e)} = \begin{bmatrix}
		\bar{\M} \dot{\H} \diag\{\x_1^{(e)}\} & & & \\
		& & \ddots&  \\
		& & & \bar{\M} \dot{\H } \diag\{\x_{K_p}^{(e)}\}
	\end{bmatrix}, \\
	&\bep = [ \bep_1^T,...,\bep_{K_p}^T ]^T.
\end{align}}

As a consequence, \eqref{pro_MZHeps} is cast as
\begin{align}
	&\min_{\bep}~\bep^T \R^{(e)} \bep+\bep^T \r^{(e)}  \nonumber \\
	&~~\text{s.t.}~~|\epsilon_d| \leq \Delta f/2,   d = 1,..., K_pD. \label{pro_bep}
\end{align}
where
\textcolor{black}{
\begin{align}
	&\R^{(e)}= \Re\{\tilde{\H}_x^{{(e)}^H }\tilde{\H}_x^{(e)}\}, \label{def_R_e} \\
	&\mathbf{r}^{(e)}	=-2 \Re \{ \tilde{\H}_x^{{(e)}^H }\tilde{\z}_x^{(e)}\},
\end{align}}
depend on the outer MBI iteration index $e$. The optimal solution $\bep^{(e+1)}$ and the corresponding optimal objective function value $\hat{u}_2^{(e+1)}=(K_p+K_s) \sum_{k=1}^{K_p}\|\bar{\M}\bar{\z}_k-\bar{\M}\bar{\H}_{\bep_k^{(e+1)}}\x_k^{(e)}\|^2 + \sum_{d=1}^D \sum_{k=1}^{K_p} \bar{c}_{d,k}| x_{d,k}^{(e)} |^2$ can be efficiently obtained resorting to the MBI approach \cite{10891687,11204963}.
Let $\mathcal{Q}^{(t)}=\{\epsilon_1^{(t)},...,\epsilon_{D K_p}^{(t)}\}$ represent the optimized variables at the $t$-th inner iteration \textcolor{black}{(for notational simplicity, the explicit dependence on the outer iteration index $e$ is omitted throughout).} The objective function value of \textcolor{black}{Problem} \eqref{pro_bep} evaluated at $\mathcal{Q}^{(t)}$ is denoted by $v(\mathcal{Q}^{(t)})$. Then the inner MBI approach requires solving the following $K_pD$ subproblems at each iteration, namely
\begin{align}
	&\min_{\epsilon_d}R_{d,d}^{(e)}\epsilon_d^2+\( 2\sum_{k=1,k\neq d}^{D K_p} R_{d,k}^{(e)}\epsilon_k^{(t)}+r_d^{(e)} \)\epsilon_d \nonumber \\
	&\text{s.t.}~~|\epsilon_d| \leq \Delta f/2,
\end{align}
\textcolor{black}{for $d=1,...,DK_p$,} where $R_{d,d}^{(e)}$, $\epsilon_k^{(t)}$ and $r_d^{(e)}$ represent $(d,d)$-th entry of $\R^{(e)}$, the $k$-th element of $\bep^{(t)}$ at the $t$-th iteration, and the $d$-th element of $\r^{(e)}$, respectively.  \textcolor{black}{The closed-form solution to $\epsilon_d^{(t+1)}$ is given by}
\begin{align}
	&\epsilon_d^{t+1} = \min\(\max\(-\frac{ 2\sum \limits_{k=1,k\neq d}^{D K_p} R_{d,k}^{(e)}\epsilon_k^{(t)}+r_d^{(e)} }{2R_{d,d}^{(e)}},-\frac{\Delta f}{2}\)  ,\frac{\Delta f}{2}\), \label{def_eps_d}
\end{align}
and $\tilde{v}_d^{(t)},d=1,...,DK_p$, is the corresponding optimal objective function value.
	 The block yielding the maximum improvement is selected according to $d^{\star}=\arg \min  \limits_{d=1,...,DK_p} \tilde{v}_d^{(t)}$ and update the $d^{\star}$-th block in $\mathcal{Q}^{(t)}$ to obtain $\mathcal{Q}^{(t+1)}$. 

Starting from $\bep^{(t=0)}=\0$, the above procedure is repeated until
$|v(\mathcal{Q}^{(t)})-v(\mathcal{Q}^{(t-1)}) |  \leq 10^{-6}$ or $t > t_{max}$, where $t_{max}$ denotes the maximum number of inner MBI procedure. The resulting solution is denoted by
$\bep^{(e+1)}=[\epsilon_1^{(t)},\ldots,\epsilon_{DK_p}^{(t)}]^T$.

\textcolor{black}{\textit{Remark:} Problem \eqref{pro_MZHeps_X} can be further modified by adding an $l_2$-norm penalty $\eta \|\bep\|^2$ to promote the estimates of the nonzero entries in the vector $\bep$ \cite{10891687}. This penalty only slightly alters the objective function in Problem \eqref{pro_bep}, while making Problem \eqref{pro_bep} strictly convex for a small positive value of $\eta$. In this work, $\eta=10^{-5}$ is adopted.}

 
 Hence, the ORDP-off algorithm employing MBI procedure is summarized in \textbf{Algorithm} \ref{RDP-OF}.

\begin{algorithm}[h]
	\caption{ ORDP-off algorithm}
	\label{RDP-OF}
	\KwIn{ \textcolor{black}{$\bar{\Z}$}, $\Z_s$\textcolor{black}{,} $\H$, $\gamma$, $q \in (0,1)$, $t_{\text{max}}$, $e_{\text{max}}$ and $\xi$.}
	\textbf{Initialization:} Set $e=0$, $\X^{(0)}= \diag(\| \H(:,1)\|,...,\| \H(:,D)\|)^{-2} \H^H\bar{\Z}$, and $\bep^{(0)}=\0$. \\
	\Repeat{$|u(\mathcal{Y}^{(e)})-u(\mathcal{Y}^{(e-1)} ) |  \leq \xi$ or $e > e_{\text{max}}$}{
		Assign $\bar{\X}=\X^{(e)}$ and $\bar{\bep}=\bep^{(e)}$. \\
		Compute $\bar{\S}(\bar{\X},\{\bar{\bep}_k\}_{k=1}^{K_p} ) = \sum_{k=1}^{K_p}(\bar{\z}_k-\bar{\H}(\bar{\bep}_k)\bar{\x}_k)(\bar{\z}_k-\bar{\H}(\bar{\bep}_k)\bar{\x}_k)^H+\Z_s  \Z_s^H$.\\
		Compute $\bar{c}_{d,k} = (| \bar{x}_{d,k} |^2+\gamma)^{\frac{q}{2}-1}$,$d=1,...,D,k=1,...,K_p$. Assign $\C_x = \diag\{ \bar{c}_{1,1},\bar{c}_{2,1},...,\bar{c}_{D,K_p} \}$. \\
		Compute $\tilde{\z}$ and $\tilde{\H}^{(e)}_{\epsilon}$ based on \eqref{def_tz_off} and \eqref{def_tH_off}, respectively.\\
		Compute $\x^{(e+1)}	=\({\tilde{\H}_{\epsilon}{^{(e)^H}}} \tilde{\H}^{(e)}_{\epsilon}  +  \frac{\C_x}{(K_p+K_s)}\)^{-1}  {\tilde{\H}_{\epsilon}{^{(e)^H}}}\tilde{\z}$, and evaluate  $\hat{u}_1^{(e+1)}$.\\
		Determine $\bep^{(e+1)}$ by via the inner MBI procedure based on \eqref{def_R_e}-\eqref{def_eps_d} and evaluate $\hat{u}_2^{(e+1)}$; \\
		Select $k^{\star}=\arg \min \limits_{k=1,2}\hat{u}_k^{(e+1)}$, and update the block $k^{\star}$ in $\mathcal{Y}^{(e)}$. \\
		Set $e=e+1$.}
	\KwOut{\textcolor{black}{Estimated RDP $\hat{\X}=\X^{(e)}$ and frequency displacement vector $\hat{\bep} = \bep^{(e)}$. }}
\end{algorithm}

\subsection{TRDP-off algorithm}

In this case, the SCM $\hat{\M}=\frac{1}{K_s} \Z_s\Z_s^H$ is utilized as the interference-plus-noise covariance matrix. As in \eqref{min_XZHX_S}, the problem for estimating $\{\bep_k\}_{k=1}^{K_p}$ and recovering $\X$ \textcolor{black}{under the} off-grid case \textcolor{black}{umbrella is} formulated as
\begin{align}
	&\min_{\X,\{\bep_k\}_{k=1}^{K_p}}~\sum_{k=1}^{K_p} (\bar{\z}_k-\bar{\H}_{\bep_k}\x_k)^H \hat{\M}^{-1} (\bar{\z}_k-\bar{\H}_{\bep_k}\x_k) \nonumber \\
	&~~~~~~~~~~~~~~~~~~~~+f_1(\X) \nonumber \\	 
	&~~\text{s.t.}~~|\epsilon_{d,k}| \leq \Delta f/2,   d = 1,...,D; k=1,...,K_p. \label{Pro_X_M_OFU}
\end{align}

Let $\check{\M}$ \textcolor{black}{denote} a matrix satisfying $\hat{\M}^{-1} = \check{\M}^H \check{\M}$. Then, with the surrogate function of $f_1(\X)$ in \eqref{major_f1}, the surrogate optimization problem associated with \eqref{Pro_X_M_OFU} is
\textcolor{black}{
\begin{align}
		&\min_{\X,\{ \bep_k \}_{k=1}^{K_p}}~ \sum_{k=1}^{K_p}\|\check{\M}\bar{\z}_k-\check{\M}\bar{\H}_{\bep_k}\x_k\|^2 + \sum_{d=1}^D \sum_{k=1}^{K_p} \bar{c}_{d,k}| x_{d,k} |^2,\nonumber \\
		&~~\text{s.t.}~~|\epsilon_{d,k}| \leq \Delta f/2,   d = 1,...,D; k=1,...,K_p. \label{pro_SCMZHeps_X}
\end{align}}

Here, \textcolor{black}{Problem} \eqref{pro_SCMZHeps_X} admits the same \textcolor{black}{structure} as \eqref{pro_MZHeps_X}, therefore, the solution details are omitted for brevity.

\section{Simulation Results}

The purpose of this section is to assess the \textcolor{black}{effectiveness} of the proposed algorithms to perform reliable RDP recovery, in terms of scatterer \textcolor{black}{parameters} estimation, miss and false detections. The developed strategies, namely ORDP-on, TRDP-on, ORDP-off, and TRDP-off, are compared with the iterative adaptive approach (IAA) \cite{5417172} and an SCM-loading two-step IAA, referred to as TIAA, where the SCM is incorporated as a covariance-loading term into the standard IAA procedure. Furthermore, the algorithms equipped with the BIC-based refinement strategy are indicated with the suffix ``-BIC''. For example, the ORDP-on with the BIC-based refinement is referred to as ORDP-on-BIC.

The considered sensing system operates at a carrier frequency $f_0 = 9.6$GHz with a range resolution of 3m \cite{10371184}. The signal model comprises $N=64$ pulses and pulse repetition interval $T=1250$ $\mu s$ \cite{collier2014polarimetric}. As a consequence, the Doppler frequency resolution is $1/(NT) = 12.5$ Hz and the frequency bin spacing is set to $\Delta f = 6.25$ Hz, i.e., half the Doppler resolution, resulting in $D=128$ Doppler bins in the unambiguous interval $\[0,800\]$Hz. Moreover, the disturbance covariance matrix is given by $\M=\M_c+\sigma_n^2\I_N$, where the $(m,n)$-th entry of $\M_c$ is \cite{11081808}
	\begin{align}
		\M_c(m,n) = \sigma_i^2 \rho^{|m-n|},
	\end{align}
with $\rho=0.9$ is the one-lag correlation coefficient. Unless otherwise specified, $\sigma_i^2=30$dB and $\sigma_n^2=0$dB. For $i$-th scatterer, with mean power $\sigma_{T,i}^2$, its integrated signal-to-interference-plus-noise ratio (SINR) is given by
	\begin{align}
		\text{SINR}_i = \sigma_{T,i}^2 \h(\tilde{f}_i)^H \M^{-1} \h(\tilde{f}_i).
	\end{align}
Subsections VI.A and VI.B deal with on-grid and off-grid cases, respectively.

\subsection{On-grid RDP Recovery}

\begin{figure}[htbp]
	\centering
	\subfloat[]{\label{RDPX0}\includegraphics[width=0.5\columnwidth]{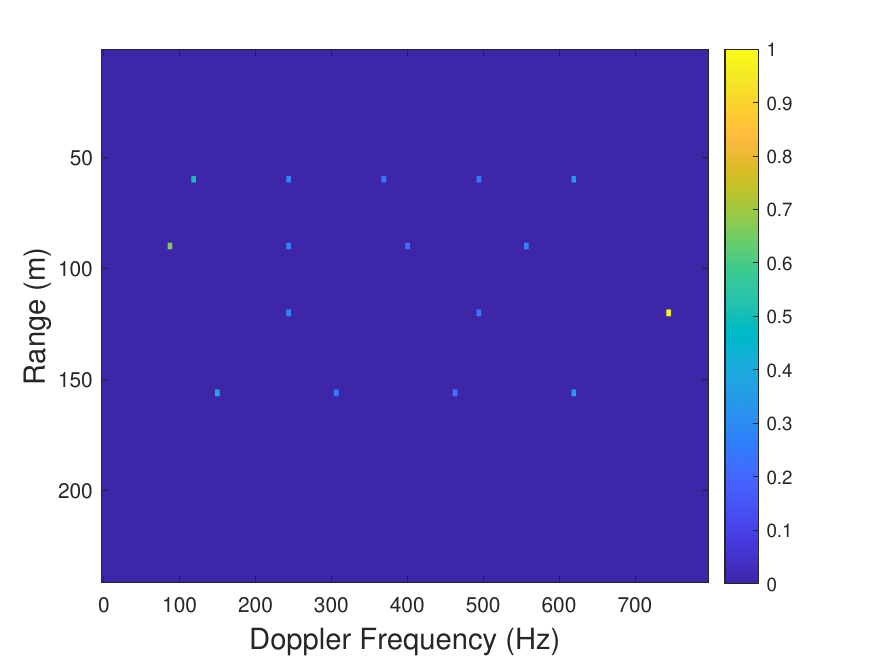} }  
	\subfloat[]{\label{IAA_BIC_SINR18}\includegraphics[width=0.5\columnwidth]{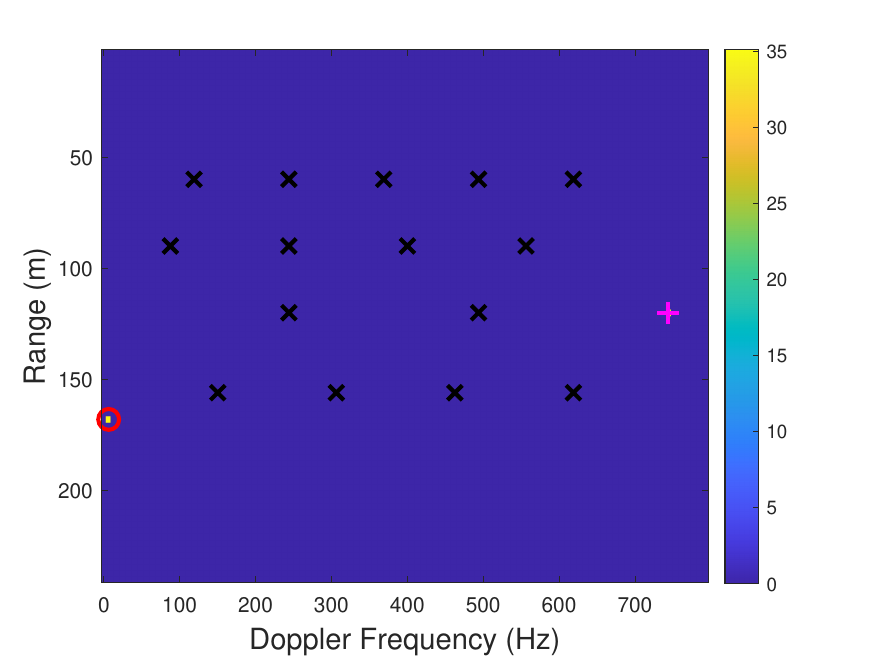} }  \newline
	\subfloat[]{\label{TIAA_BIC_SINR18}\includegraphics[width=0.5\columnwidth]{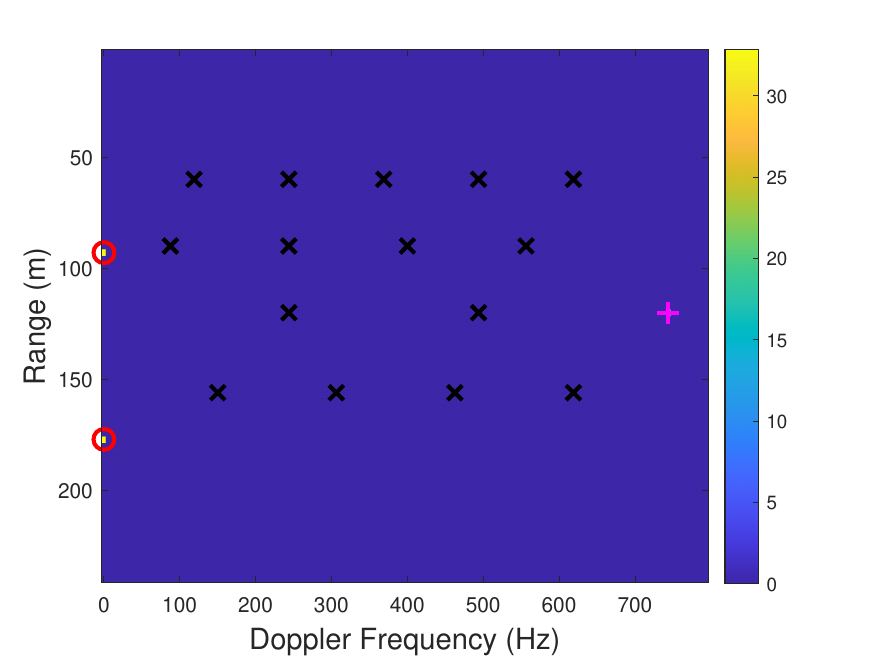} } 
	\subfloat[]{\label{TRDP_on_SINR18_bic}\includegraphics[width=0.5\columnwidth]{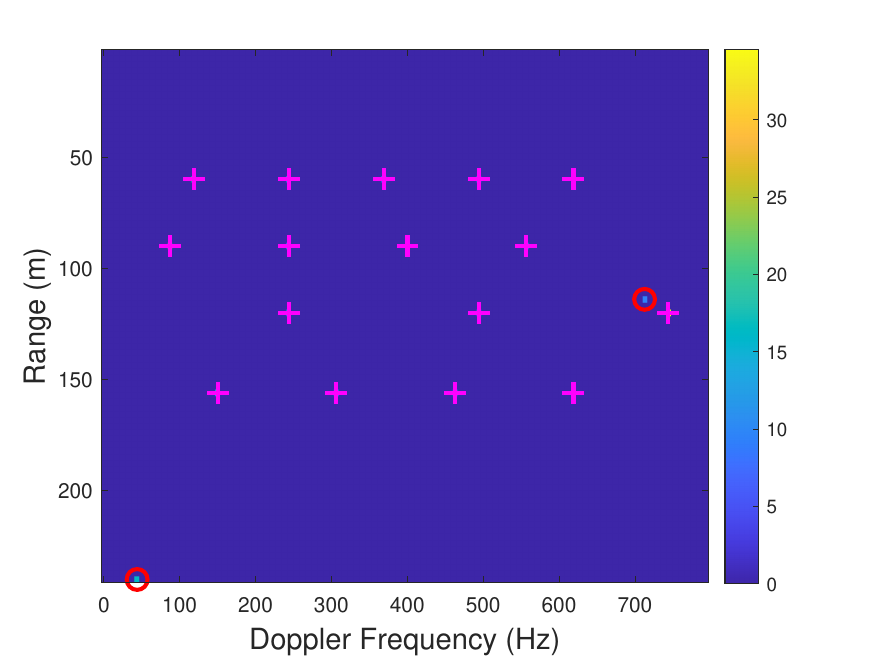} } \newline
	\subfloat[]{\label{ORDP_on_SINR18_bic}\includegraphics[width=0.5\columnwidth]{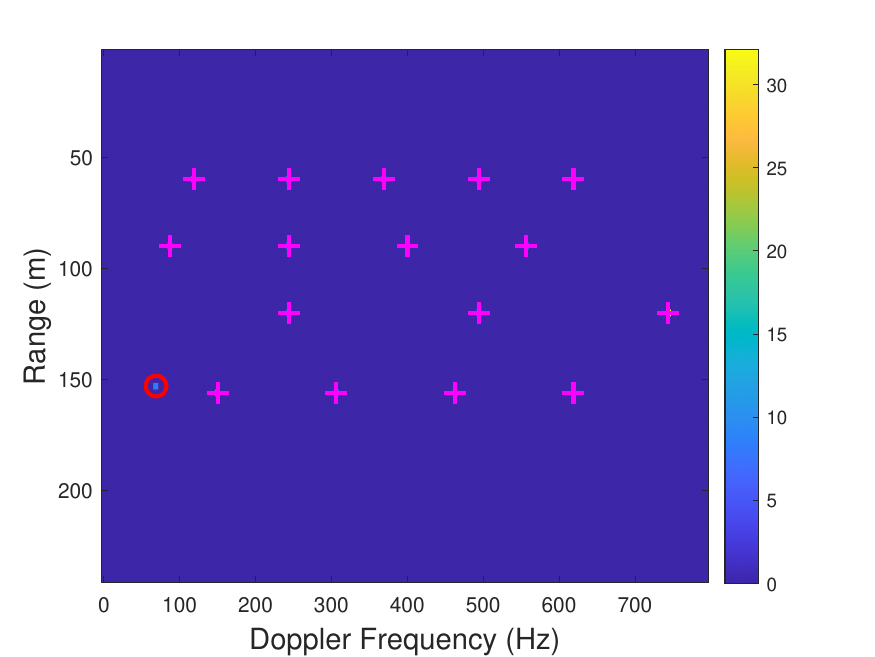} }\newline
	\caption{(a) shows true RDP; (b), (c), \textcolor{black}{(d), and (e)} are obtained RDPs by IAA-BIC, TIAA-BIC, \textcolor{black}{TRDP-on-BIC, and} ORDP-on-BIC methods, respectively, \textcolor{black}{assuming SINR=$18$dB and $K_s=200$}.}
	\label{RDP_SINR_18dB_IAA}
\end{figure}

In this subsection, several experiments are conducted to evaluate RDP recovery performance for the on-grid case. Here, the region of interest is \textcolor{black}{composed of} $K_p$=80 range bins, corresponding to a 240m range swath. 16 scatterers with identical signal-to-interference-plus-noise ratio (SINR) are assumed. Their range-Doppler coordinates (expressed in $[\mathrm{m},\mathrm{Hz}]$) are $[60,118.75]$, $[60,243.75]$, $[60,368.75]$, $[60,493.75]$, $[60,618.75]$, $[90,87.50]$, $[90,243.75]$, $[90,400.00]$, $[90,556.25]$, $[120,243.75]$, $[120,493.75]$, $[120,743.75]$, $[156,150.00]$, $[156,306.25]$, $[156,462.50]$, and $[156,618.75]$\textcolor{black}{, and the resulting} normalized RDP map $\X_0$ is shown in Fig. \ref{RDP_SINR_18dB_IAA}\subref{RDPX0}.

To evaluate the quality of the recovered RDP, the estimate of the normalized mean squared error (NMSE) between the recovered and true RDP is computed via \textcolor{black}{$L=100$} independent Monte Carlo (MC) trials, \textcolor{black}{hereafter referred to as E-NMSE-X}; specifically, let $\hat{\X}_l$ denote the RDP estimated in the $l$-th MC run using a given recovery technique, then
\begin{align}
	\text{E-NMSE-X}=\frac{1}{L}	\sum_{l=1}^{L}\frac{\| \X_0 - \hat{\X}_l \|^2 }{ \| \X_0\|^2 }.
\end{align}

Regarding the parameter employed by ORDP-on, TRDP-on, ORDP-on-BIC and TRDP-on-BIC algorithms, they are chosen as $\gamma=10^{-6}$, $e_{\text{max}}=2000$, \textcolor{black}{$t_{\text{max}} = 1000$}, $q=0.6$, and $K_s=200$. The convergence tolerances for the RML optimization and BIC refinement are $\xi=10^{-4}$ and $\xi=10^{-6}$, respectively. 

\textcolor{black}{In Figs. \ref{RDP_SINR_18dB_IAA}(b)-\ref{RDP_SINR_18dB_IAA}(e), the results of RDP recovery process provided by IAA-BIC, TIAA-BIC, TRDP-on-BIC, and ORDP-on-BIC at SINR=$18$dB are illustrated, with reference to one simulation trial; therein, the magenta ``+” markers denote successfully detected true scatterers, the black ``x” markers indicate missed scatterers, and the red circles represent falsely detected scatterers.}

A visual comparison between the reconstructed RDPs and the true RDP shows that our proposed \textcolor{black}{ORDP-on-BIC and TRDP-on-BIC} methods are capable of correctly locating the true scatterers. Moreover, it can observed that the \textcolor{black}{IAA-BIC and TIAA-BIC} methods in Figs. \ref{RDP_SINR_18dB_IAA}\subref{IAA_BIC_SINR18}-\ref{RDP_SINR_18dB_IAA}\subref{TIAA_BIC_SINR18} correctly detect fewer scatterers than the proposed \textcolor{black}{TRDP-on-BIC and ORDP-on-BIC} methods in Figs. \ref{RDP_SINR_18dB_IAA}\subref{TRDP_on_SINR18_bic}-\ref{RDP_SINR_18dB_IAA}\subref{ORDP_on_SINR18_bic}. Compared with the \textcolor{black}{IAA-BIC}, the \textcolor{black}{TIAA-BIC} increases the number of false detections. Interestingly, although the proposed \textcolor{black}{ORDP-on-BIC and TRDP-on-BIC} methods achieve near-perfect recovery performance, the former generates fewer false alarms than the latter, \textcolor{black}{in this specific simulation}
\footnote{\textcolor{black}{Due to space limitations, the RDPs obtained without BIC-based refinement are not reported. Nevertheless, it is worth pointing out that the proposed refinement strategy reduces undesired responses in scatterer-free regions, resulting in visually cleaner RDPs.}}. 

\begin{figure}[htbp]
	\centering
	\subfloat[]{\label{NMSE_SINR_on}\includegraphics[width=0.5\columnwidth]{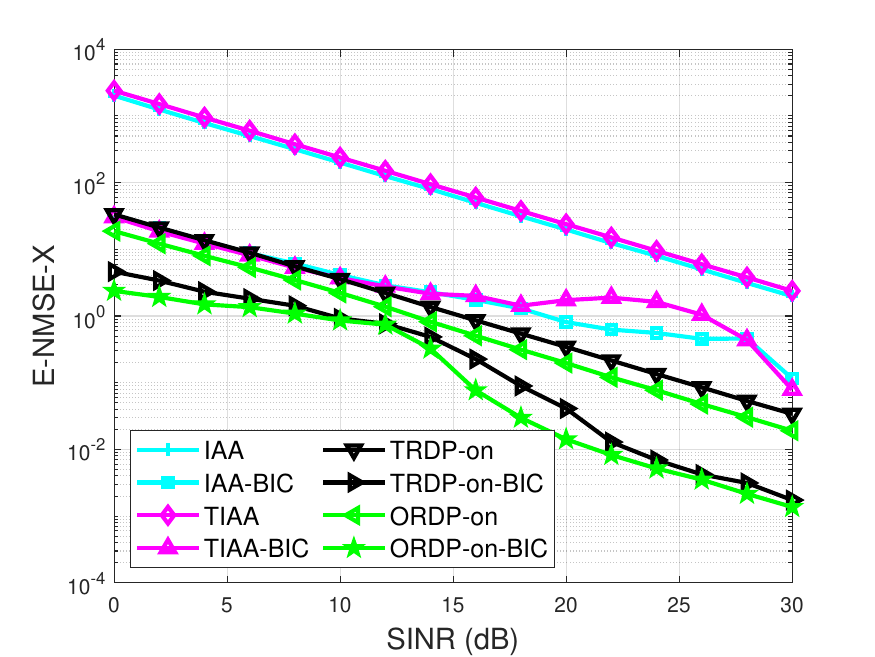} } \hspace{-0.5cm}	\subfloat[]{\label{PD_SINR_on}\includegraphics[width=0.5\columnwidth]{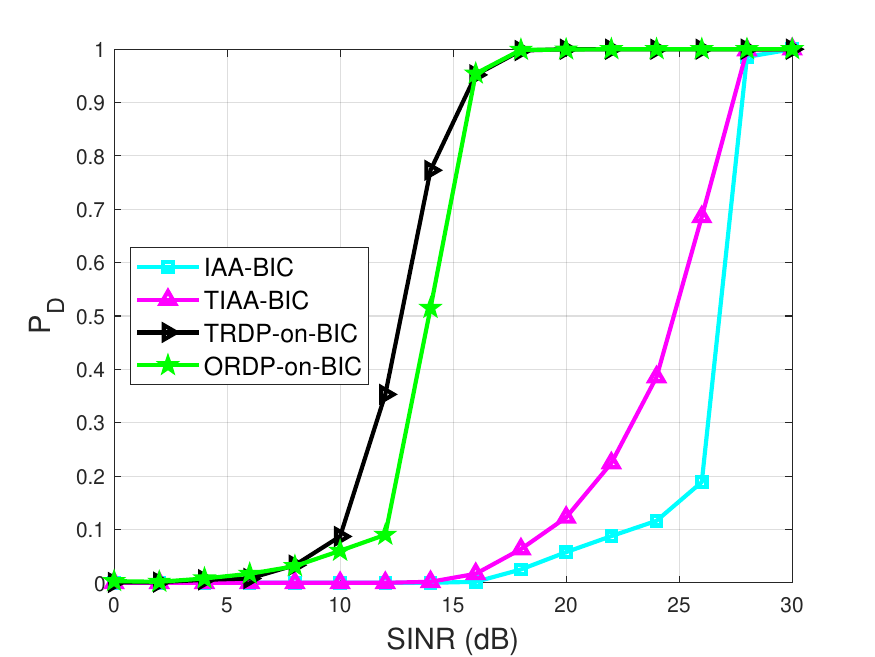} } \newline
	\subfloat[]{\label{PFA_SINR_on}\includegraphics[width=0.5\columnwidth]{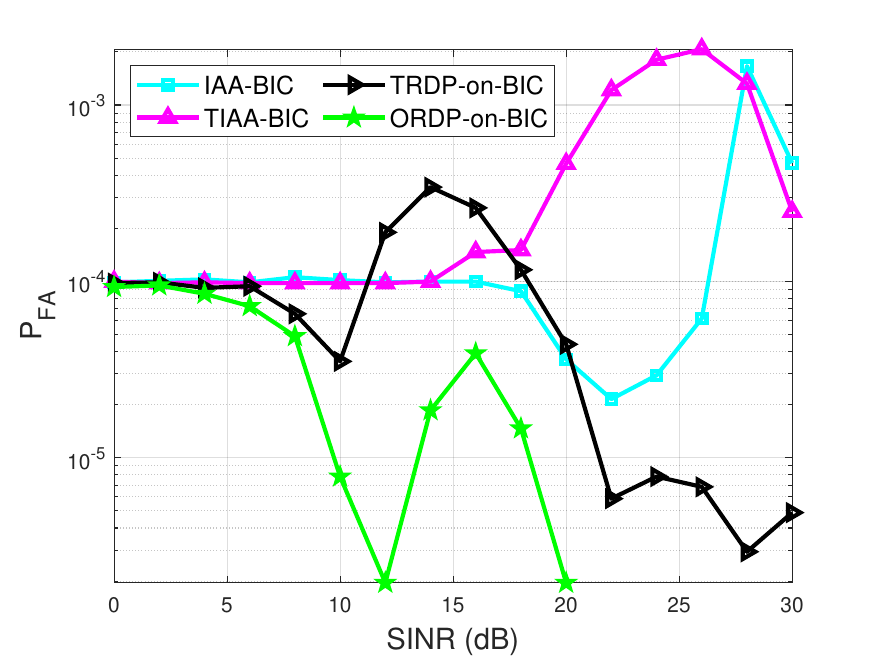} } \hspace{-0.5cm}
	\subfloat[]{\label{NMSE_Ks}\includegraphics[width=0.5\columnwidth]{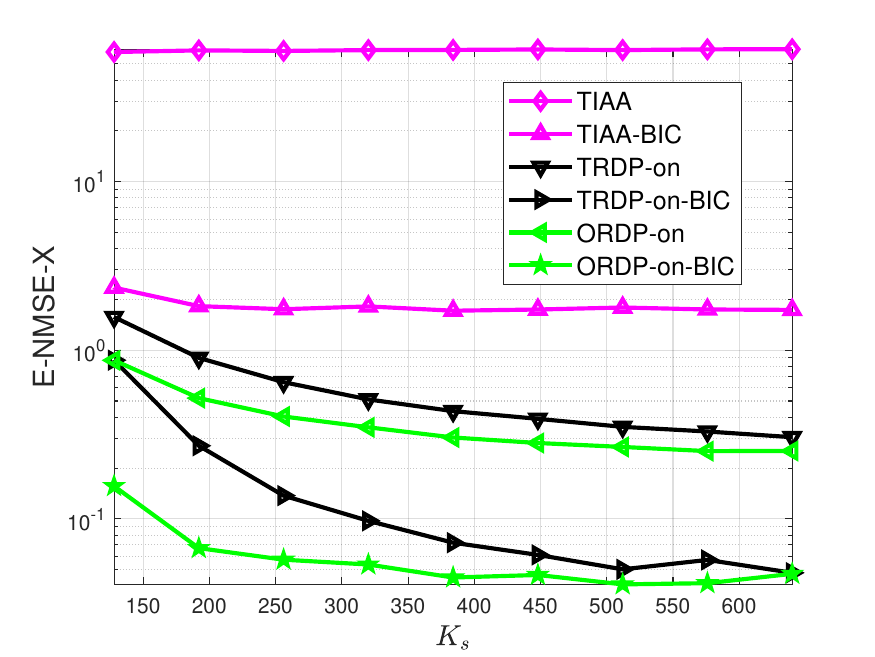} }  \newline
	\subfloat[]{\label{Pd_Ks}\includegraphics[width=0.5\columnwidth]{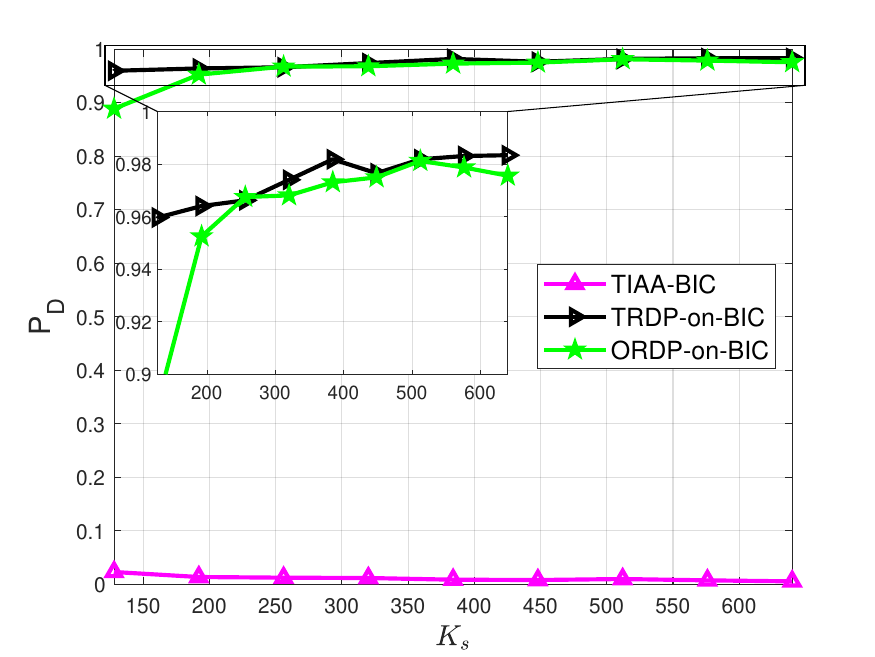} }  \hspace{-0.5cm}
	\subfloat[]{\label{PFA_Ks}\includegraphics[width=0.5\columnwidth]{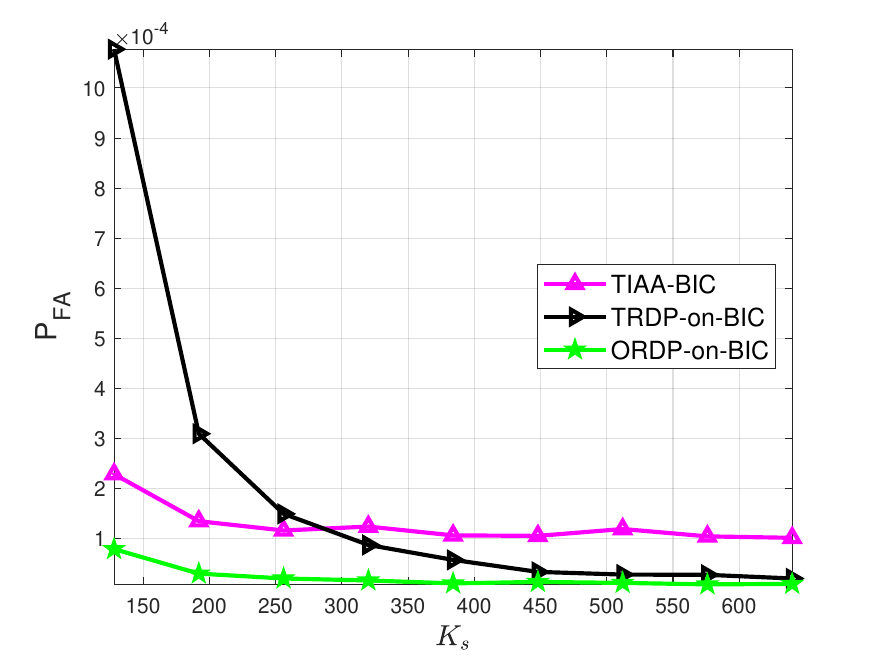} }   \newline
	\caption{(a), \textcolor{black}{(b),} and (c) are E-NMSE-Xs, \textcolor{black}{$\text{P}_\text{D}$s, and} $\text{P}_\text{FA}$s versus SINR, respectively\textcolor{black}{, assuming $K_s=200$}; (d), (e)\textcolor{black}{,} and (f) are E-NMSE-Xs, \textcolor{black}{$\text{P}_\text{D}$s, and} $\text{P}_\text{FA}$s versus $K_s$, respectively\textcolor{black}{, assuming SINR=16dB}.}
	\label{Tri_SINR_on_grid}
\end{figure}

To gain further insight and provide an overall picture of the recovery capabilities of the proposed algorithms, \textcolor{black}{the estimation accuracy and the detection performance of the different strategies are} analyzed over a range of SINR values. Figs. \ref{Tri_SINR_on_grid}\subref{NMSE_SINR_on}, \ref{Tri_SINR_on_grid}\subref{PD_SINR_on}, and \ref{Tri_SINR_on_grid}\subref{PFA_SINR_on} report the E-NMSE-X, empirical probability of detection ($\text{P}_\text{D}$) and empirical probability of false alarm ($\text{P}_\text{FA}$) versus SINR\footnote{$\text{P}_\text{FA}$ is estimated by counting detections occurring in range–Doppler bins where no scatterers are present, whereas $\text{P}_\text{D}$ is computed by counting detections in the bins corresponding to the true scatterers \cite{11204963,8574991}. Over 100 MC trials, a total of 1,022,400 scatterer-free range–Doppler bins and 1,600 scatterer-occupied bins are collected for statistical evaluation.}, respectively. Inspection of the curves shows that the proposed TRDP-on-based and ORDP-on-based algorithms outperform their IAA-based and TIAA-based counterparts. Indeed, the TRDP-on-based and ORDP-on-based algorithms achieve lower E-NMSE-X values and, particularly in the moderate-to-high-SINR region, significantly higher detection probabilities and lower false-alarm rates. These results corroborate the effectiveness of explicitly exploiting the sparsity of the RDP through an element-sparsity-promoting penalty function, thereby enhancing reconstruction performance. Moreover, the refinement process generally leads to performance improvements. Furthermore, \textcolor{black}{each} ORDP-on-based \textcolor{black}{strategy exhibits} lower E-NMSE-X than the TRDP-on-based \textcolor{black}{counterpart}, which substantiates the benefit of estimating the interference-plus-noise covariance matrix \textcolor{black}{suitably using all the data,} instead of relying on the SCM as a direct proxy. Nevertheless, in the moderate-SINR regime, TRDP-on-BIC attains a higher $\text{P}_\text{D}$ than ORDP-on-BIC, albeit at the expense of a greater $\text{P}_\text{FA}$.	
Finally, it is worth noting that  ORDP-on-based and TRDP-on-based strategies show a rapid decrease in $\text{P}_\text{FA}$ as the SINR increases (for the considered number of MC trials, no false alarms are observed for ORDP-on-BIC at SINR above 20 dB). As the SINR increases, the target amplitudes are estimated with increasing accuracy, leading to \textcolor{black}{progressively} weaker residual responses. Consequently, the BIC-based model selection procedure can yield a more accurate \textcolor{black}{determination} of the target support, containing fewer undesired residual responses in the no-scatterer region and leading to a lower empirical $\text{P}_\text{FA}$.

To further investigate the robustness and recovery capabilities of the proposed RDP recovery approaches, the second experiment examines the effect of the number of secondary data samples, $K_s$, on both recovery accuracy and detection performance. The SINR is fixed at 16 dB, whereas all other parameters remain unchanged from the previous experiment. The parameter $K_s$ ranges from 128 to 640 \textcolor{black}{in} increments of 64.

\textcolor{black}{Figs. \ref{Tri_SINR_on_grid}\subref{NMSE_Ks}, \ref{Tri_SINR_on_grid}\subref{Pd_Ks}, and \ref{Tri_SINR_on_grid}\subref{PFA_Ks} show the E-NMSE-X, the empirical $\text{P}_\text{D}$, and the empirical $\text{P}_\text{FA}$ versus $K_s$, respectively. Looking at Fig. \ref{Tri_SINR_on_grid}\subref{NMSE_Ks}, it follows \textcolor{black}{that} the proposed processing strategies outperform the TIAA-based approaches. Moreover, as $K_s$ increases, the E-NMSE-X achieved by the ORDP-on-based and TRDP-on-based methods generally decreases, indicating that the availability of additional secondary data improves the accuracy of RDP recovery. In contrast, the TIAA-based methods exhibit limited sensitivity to variations in $K_s$. Furthermore, over the entire range of $K_s$ values, the ORDP-on-based strategies consistently achieve lower E-NMSE-X values than their TRDP-on-based alternatives. This performance gap demonstrates the effectiveness of jointly inferring the RDP and the interference-plus-noise covariance matrix, rather than directly employing the SCM as an estimate of the covariance matrix. The BIC-based refinement further reduces the reconstruction error. Among all the considered algorithms, ORDP-on-BIC achieves the lowest E-NMSE-X, thereby confirming its superior RDP recovery performance.}

As shown in Fig. \ref{Tri_SINR_on_grid}\subref{Pd_Ks}, the TRDP-on-BIC and ORDP-on-BIC methods achieve high detection probabilities over the entire range of $K_s$ values. Furthermore, the $\text{P}_\text{D}$ values of TRDP-on-BIC and ORDP-on-BIC increase and saturate \textcolor{black}{fastly} to a value close to one. As to the $\text{P}_\text{FA}$ values reported in Fig. \ref{Tri_SINR_on_grid}\subref{PFA_Ks}, they highlight that ORDP-on-BIC consistently yields the lowest $\text{P}_\text{FA}$, whereas the $\text{P}_\text{FA}$ of TRDP-on-BIC decreases rapidly with $K_s$ and falls below that of TIAA-BIC when $K_s$ exceeds 320. Additionally, the  $\text{P}_\text{FA}$ values of the ORDP-on-BIC and TRDP-on-BIC  generally decrease and stabilize as $K_s$ increases, demonstrating that additional secondary data \textcolor{black}{also} contributes to a reduction in false alarms.

\subsection{Off-grid RDP Recovery}

In this subsection, several experiments are conducted to evaluate RDP recovery and Doppler frequency displacement estimation performance in the off-grid scenario. Compared with the on-grid case, the region of interest is restricted to a smaller range interval in order to focus on the scatterer-populated area. Specifically, the 16 scatterers with identical SINR described in Section VI-A are retained, while the observation range is narrowed from [0,240]m with $K_p=80$ range bins to [48,168]m with $K_p=40$ range bins. For the 16 scatterers specified in Section VI-A, the Doppler frequency displacements, arranged in the same order as \textcolor{black}{the previously} listed range-Doppler \textcolor{black}{pairs}, are set to $\bep_0=$[-0.92, 0.75, -1.19, 0.89, -0.76, -0.9, -0.8, 0.97, -1.18, -1.20, 1.4, -1.29, 1.20, 1.2, -0.87, 0.95]$^T$Hz.

To measure the accuracy of the frequency displacement estimation using the proposed strategies, the \textcolor{black}{E-NMSE} between the estimated and true frequency displacement is computed via $L=100$ MC trials, \textcolor{black}{denoted henceforth by} E-NMSE-$\epsilon$.
 \textcolor{black}{To this end, the focus is only on the bins occupied by the scatterers \cite{10891687}.}
 Specifically, let $\hat{\bep}_l$ denote the frequency displacement estimated in the $l$-th MC run using a given recovery strategy, then
 \textcolor{black}{
\begin{align}\label{def_E_NMSE_X}
	\text{E-NMSE-}\epsilon=\frac{1}{L}	\sum_{l=1}^{L}\frac{\| \bep_0 - \hat{\bep}_l \|^2 }{ \text{Card}\{ \bep_0 \} \Delta f^2 }.
\end{align}}

\begin{figure}[htbp]
	\centering
	\subfloat[]{\label{NMSE_SINR_off}\includegraphics[width=0.5\columnwidth]{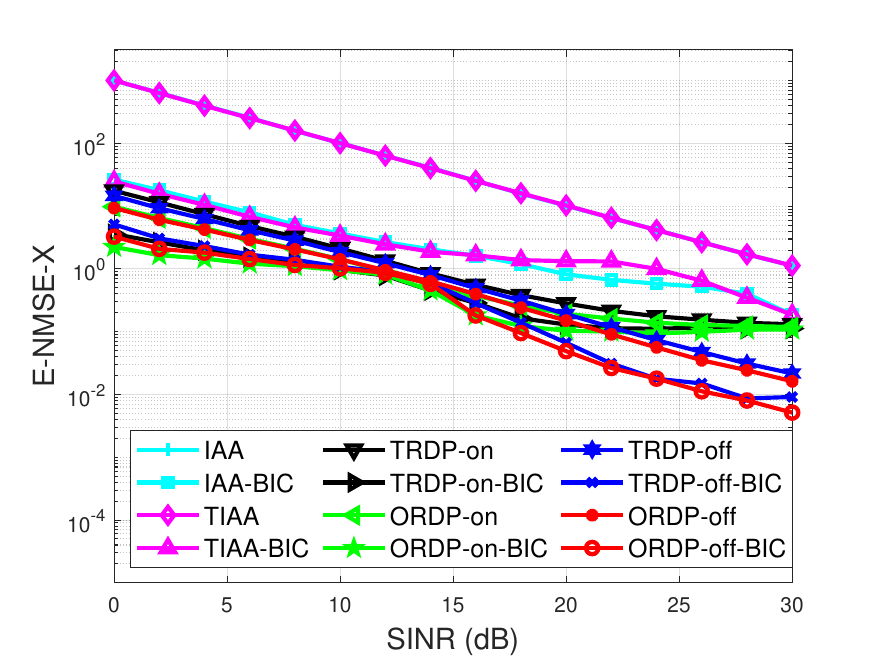} }  \hspace{-0.5cm}
	\subfloat[]{\label{NMSE_Dtheta_SINR_off}\includegraphics[width=0.5\columnwidth]{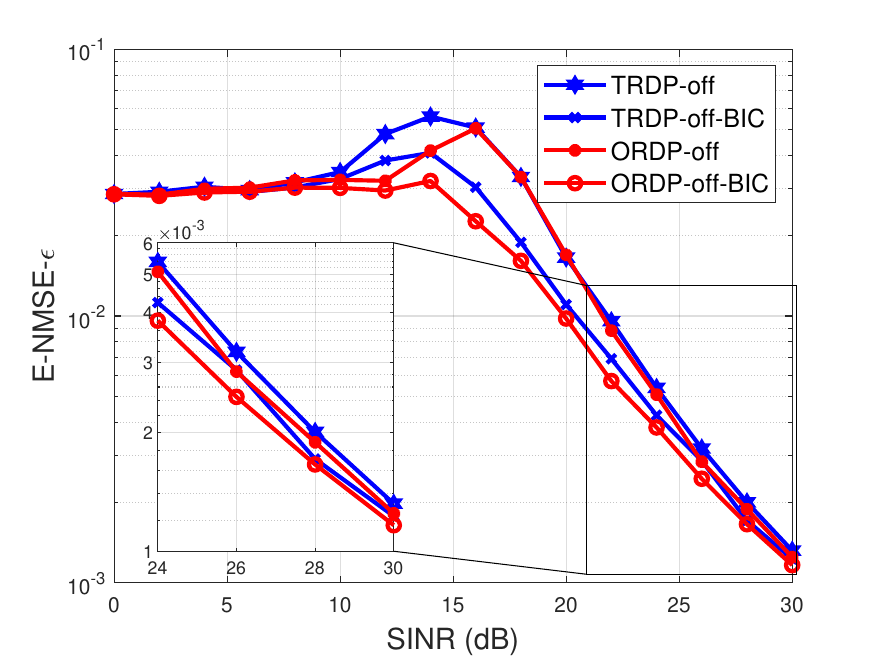} } \newline
	\subfloat[]{\label{Pd_SINR_off}\includegraphics[width=0.5\columnwidth]{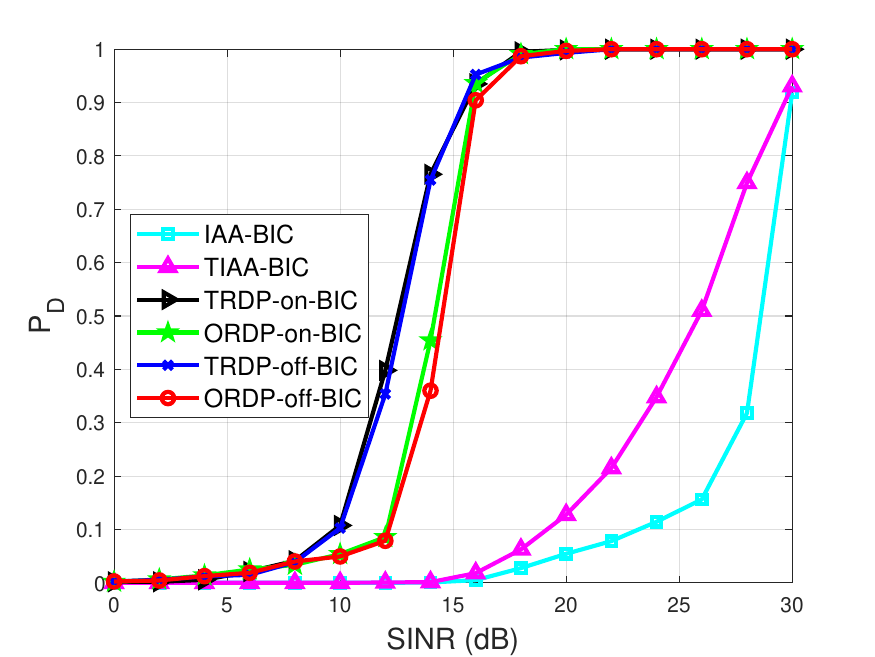} } \hspace{-0.5cm}
	\subfloat[]{\label{PFA_SINR_off}\includegraphics[width=0.5\columnwidth]{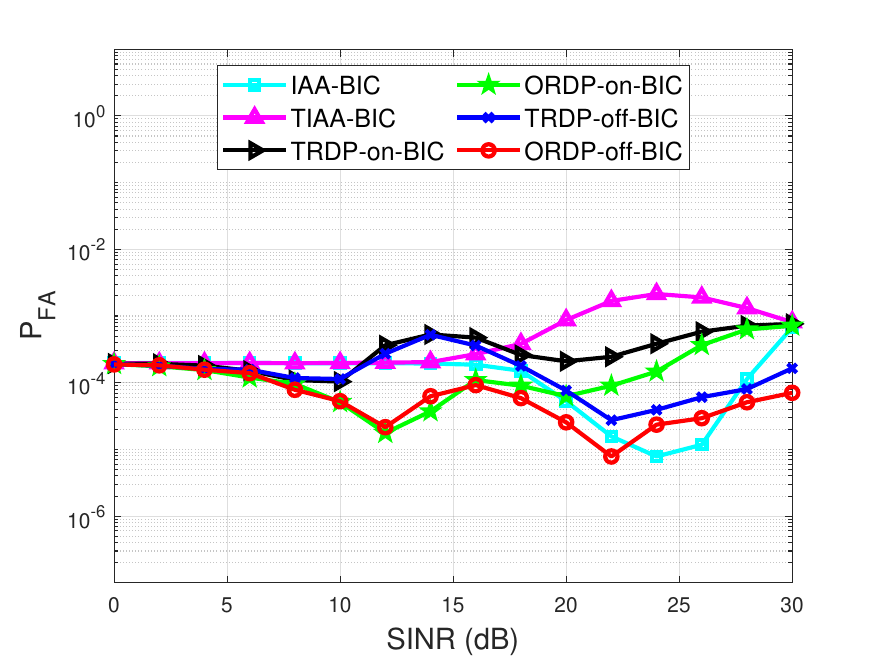} } \newline
	\caption{(a), (b), \textcolor{black}{(c),} and (d) are E-NMSE-Xs, E-NMSE-$\epsilon$s, \textcolor{black}{$\text{P}_\text{D}$s, and} $\text{P}_\text{FA}$s versus SINR, respectively, \textcolor{black}{assuming $K_s=200$.}
		}
	\label{Tri_SINR_on_SINR_off}
\end{figure}

\textcolor{black}{Regarding the parameter employed by TRDP-off, ORDP-off algorithms, they are set as $\gamma=10^{-6}$, $e_{\text{max}}=2000$, \textcolor{black}{$t_{\text{max}} = 1000$}, $q=0.6$, and $K_s=200$, while the convergence tolerances for the RML optimization and BIC refinement are $\xi=10^{-4}$ and $\xi=10^{-6}$, respectively.} Figs. \ref{Tri_SINR_on_SINR_off}\subref{NMSE_SINR_off}, \ref{Tri_SINR_on_SINR_off}\subref{NMSE_Dtheta_SINR_off}, \ref{Tri_SINR_on_SINR_off}\subref{Pd_SINR_off}, and \ref{Tri_SINR_on_SINR_off}\subref{PFA_SINR_off} report the E-NMSE-Xs, E-NMSE-$\epsilon$s, \textcolor{black}{empirical $\text{P}_\text{D}$s,} and $\text{P}_\text{FA}$s versus SINR for the considered recovery strategies, including the on-grid approaches (IAA-based, TIAA-based, TRDP-on-based, and ORDP-on-based), as well as the proposed off-grid \textcolor{black}{strategies} (TRDP-off-based and ORDP-off-based).

\textcolor{black}{The curves in Fig. \ref{Tri_SINR_on_SINR_off}\subref{NMSE_SINR_off} show} that, in \textcolor{black}{the} off-grid scenario, the ORDP-on-based and TRDP-on-based strategies continue to achieve lower E-NMSE-X \textcolor{black}{values} than the IAA-based and TIAA-based counterparts, \textcolor{black}{consistent with} the results in \textcolor{black}{the} on-grid case. Moreover, in the low-moderate-SINR region, i.e., from 0dB to 16dB, the TRDP-off-based and ORDP-off-based methods exhibit comparable E-NMSE-X values to their TRDP-on-based and ORDP-on-based counterparts, respectively. As the SINR increases beyond 18dB, the advantage of accounting for the frequency displacement becomes increasingly \textcolor{black}{higher}, leading to progressive reductions in E-NMSE-X. \textcolor{black}{Indeed}, the E-NMSE-X curves associated with the TRDP-on-based and ORDP-on-based methods approach a floor when the SINR exceeds 18dB. Furthermore, the consistently lower \textcolor{black}{E-NMSE-Xs} attained by the ORDP-off-based strategies relative to \textcolor{black}{their TRDP-off-based counterparts substantiate} the benefit of jointly \textcolor{black}{exploiting primary and secondary data to estimate} the interference-plus-noise covariance matrix \textcolor{black}{and RDP also in} the off-grid case \textcolor{black}{as well.} Looking at Fig. \ref{Tri_SINR_on_SINR_off}\subref{NMSE_Dtheta_SINR_off}, the ORDP-off-based methods achieve lower E-NMSE-$\epsilon$ than their TRDP-off-based counterparts \textcolor{black}{in moderate-to-high-SINR}. In particular, ORDP-off-BIC attains the lowest E-NMSE-$\epsilon$ over the considered SINR range, demonstrating the effectiveness of \textcolor{black}{combining} joint estimation and \textcolor{black}{BIC-based} refinement. Although the performance gaps narrow as the SINR increases, the advantage of the ORDP-off-based methods remains visible in the enlarged subfigure.

As shown in Fig. \ref{Tri_SINR_on_SINR_off}\subref{Pd_SINR_off}, \textcolor{black}{as} in the on-grid case, the IAA-based and TIAA-based methods require higher SINR to attain $\text{P}_{\text{D}}$ values \textcolor{black}{comparable to} those of the proposed methods. 
\textcolor{black}{
Moreover, the improvement in $\text{P}_{\text{D}}$ offered by TRDP-off-BIC over ORDP-off-BIC comes at the cost of an increased $\text{P}_{\text{FA}}$,
}
as shown in Fig. \ref{Tri_SINR_on_SINR_off}\subref{PFA_SINR_off}. Although TRDP-off-BIC and ORDP-off-BIC achieve comparable $\text{P}_{\text{D}}$s to those of TRDP-on-BIC and ORDP-on-BIC, respectively, their false-alarm performance differs markedly as the SINR increases. \textcolor{black}{As the SINR increases, the energy leaked from the target Doppler bin into \textcolor{black}{adjacent cell} becomes more pronounced, causing some neighboring bins to be erroneously identified as true scatterers and thereby increasing the false-alarm probability.} Therefore, TRDP-on-BIC and ORDP-on-BIC exhibit higher $P_{\mathrm{FA}}$ values than those of TRDP-off-BIC and ORDP-off-BIC, respectively.


\begin{figure}[htbp]
	\centering
	\subfloat[]{\label{NMSE_Ks_off}\includegraphics[width=0.5\columnwidth]{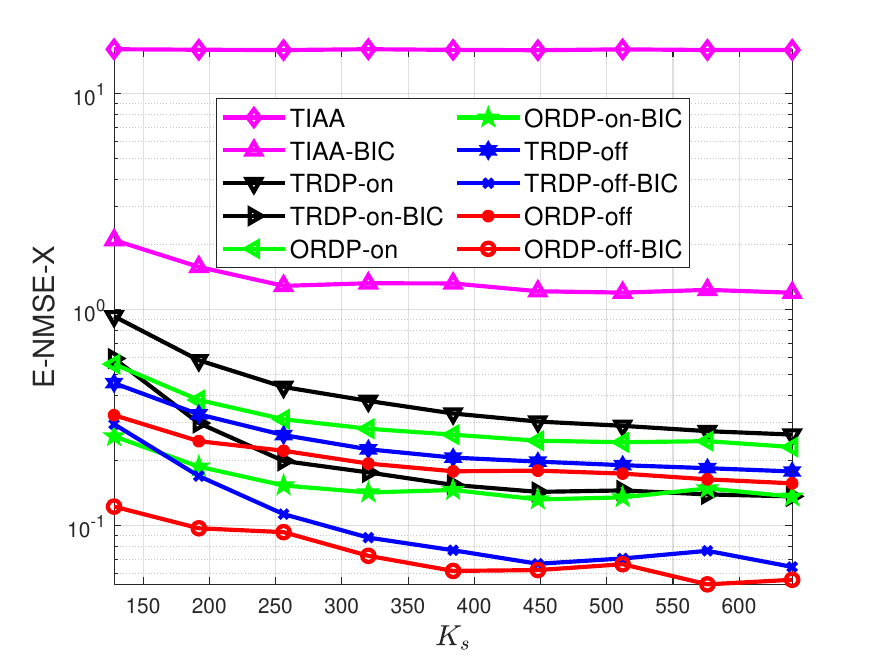} }  \hspace{-0.5cm}
	\subfloat[]{\label{NMSE_Dtheta_Ks_off}\includegraphics[width=0.5\columnwidth]{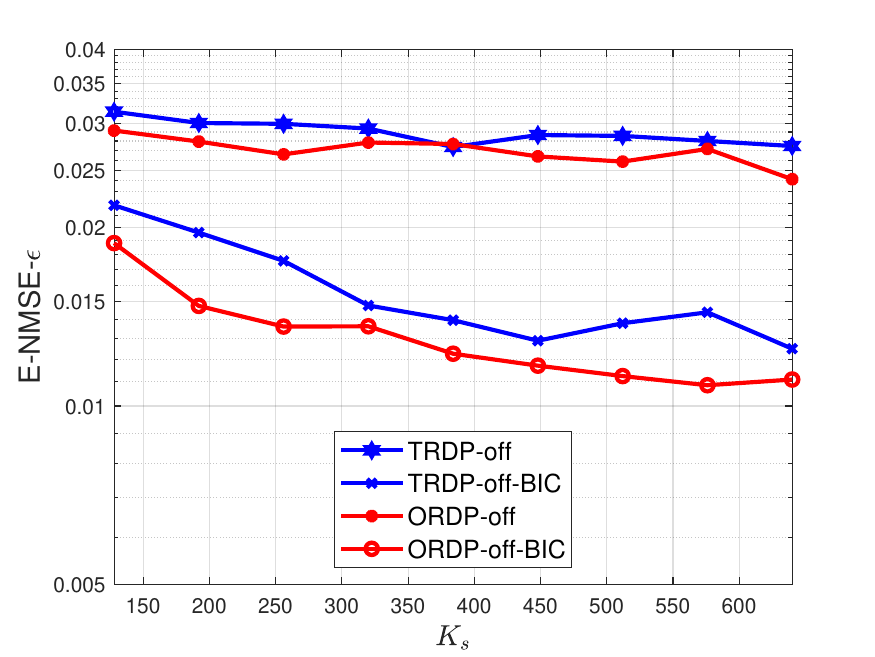} } \newline
	\subfloat[]{\label{Pd_Ks_off}\includegraphics[width=0.5\columnwidth]{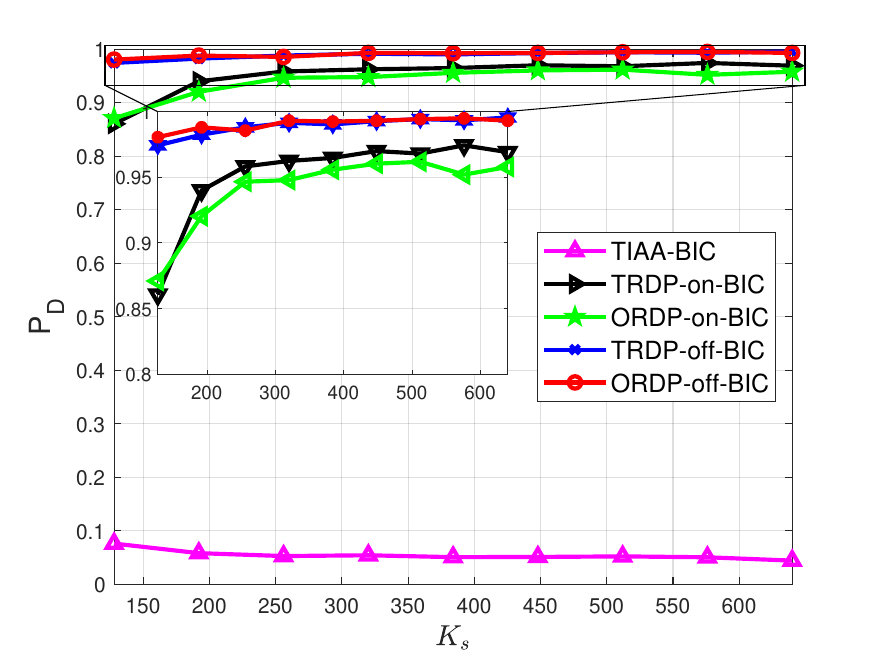} } \hspace{-0.5cm}
	\subfloat[]{\label{PFA_Ks_off}\includegraphics[width=0.5\columnwidth]{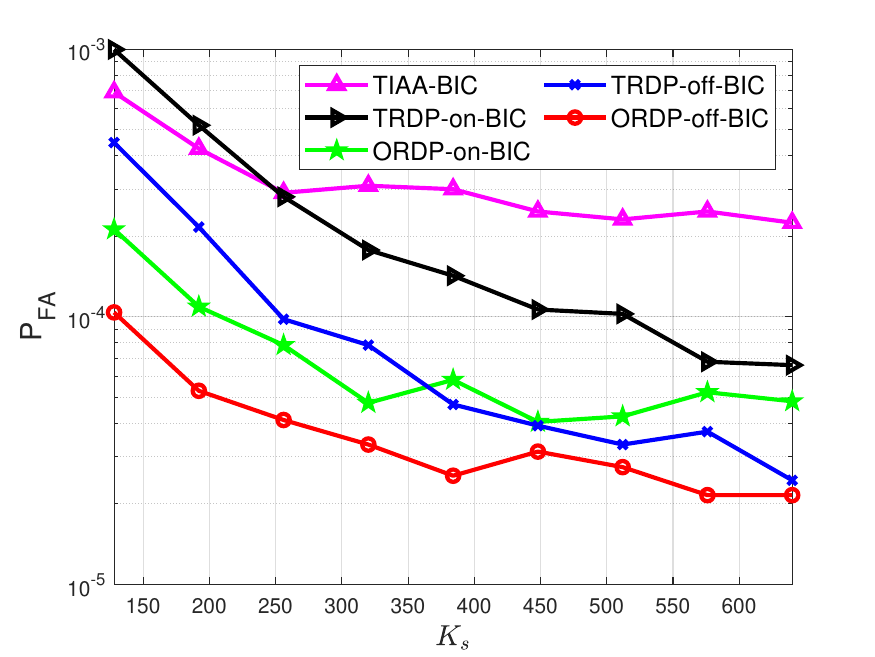} } \newline
	\caption{(a), (b), \textcolor{black}{(c), and (d)} are E-NMSE-Xs, E-NMSE-$\epsilon$s, \textcolor{black}{$\text{P}_\text{D}$s, and} $\text{P}_\text{FA}$s versus $K_s$, respectively\textcolor{black}{, assuming SINR=18dB.} 
	}
	\label{Tri_Ks_off}
\end{figure}

To elucidate the effect of the number of secondary data samples \textcolor{black}{on} the RDP recovery and frequency-displacement estimation capabilities of the proposed off-grid strategies, all the considered methods are \textcolor{black}{analyzed} for $K_s$ ranging from 128 to 640 in increments of 64. Figs. \ref{Tri_Ks_off}(a), \ref{Tri_Ks_off}(b), \textcolor{black}{\ref{Tri_Ks_off}(c),} and \ref{Tri_Ks_off}(d) report E-NMSE-Xs, E-NMSE-$\epsilon$s, \textcolor{black}{$\text{P}_\text{D}$s,} and $\text{P}_\text{FA}$s versus $K_s$ at SINR$=$18dB, respectively. \textcolor{black}{Inspection of Fig. \ref{Tri_Ks_off}(a) unveils that} the proposed off-grid strategies achieve lower E-NMSE-X values than TIAA-based and their on-grid counterparts, demonstrating the benefit of accounting for Doppler frequency displacements. Moreover, the ORDP-off-based approaches outperform the TRDP-off-based \textcolor{black}{strategies over} the entire range of $K_s$, with the BIC-based processing providing a performance boost for all the methods. \textcolor{black}{Finally, ORDP-off-BIC achieves the lowest E-NMSE-X.}

As shown in Fig. \ref{Tri_Ks_off}(b), ORDP-off-BIC consistently achieves the lowest E-NMSE-$\epsilon$ over the considered range of $K_s$. Moreover, the ORDP-off-based strategies generally attain lower E-NMSE-$\epsilon$ than their TRDP-off-based \textcolor{black}{alternatives}. The results also show that increasing the number of secondary data samples improves the frequency-displacement estimation accuracy, particularly for the BIC-refined strategies. 
As shown in Figs.~\ref{Tri_Ks_off}(c) and \ref{Tri_Ks_off}(d), increasing $K_s$ generally enhances $\mathrm{P}_{\mathrm{D}}$ of all four proposed recovery schemes while \textcolor{black}{reducing} their $\mathrm{P}_{\mathrm{FA}}$. The TRDP-off-BIC and ORDP-off-BIC consistently attain higher $\mathrm{P}_{\mathrm{D}}$ and lower $\mathrm{P}_{\mathrm{FA}}$ than their on-grid counterparts, demonstrating the benefit of explicitly compensating for Doppler frequency displacement.
Overall, ORDP-off-BIC provides the lowest $\text{P}_\text{FA}$ among the considered methods.

\section{Conclusion}

This article has investigated sparse RDP recovery for adaptive radar systems operating in colored interference with an unknown interference-plus-noise covariance. A unified secondary-data-aided RML framework has been developed for both on-grid and off-grid scenarios without imposing prescribed parametric covariance structures. The one-step formulations jointly estimate the sparse RDP, unknown covariance matrix, and possible Doppler frequency displacements, whereas two-step alternatives are developed by replacing the unknown covariance matrix with \textcolor{black}{SCM before likelihood optimization}. Tailored MM and MM-MBI algorithms are developed for the on-grid and off-grid problems, respectively, with monotonic objective descent and KKT convergence guarantees. BIC-based refinement is further incorporated to improve support identification. Numerical results show that joint covariance estimation generally outperforms the SCM-based alternatives, while off-grid modeling effectively mitigates Doppler mismatch and improves reconstruction and detection performance. Overall, the proposed framework provides a general solution to sparse RDP recovery under covariance uncertainty and Doppler mismatch without restrictive covariance parameterizations.

Future work will extend the proposed framework to multidimensional target parameter estimation, such as joint range-Doppler-angle profile recovery for multiple-input multiple-output radar systems, and evaluate its performance using measured radar data.

\bibliographystyle{IEEEtran}
\bibliography{reference}


\end{sloppypar}
\end{document}

%% file: def.tex
\usepackage{amsfonts, amsmath, mathrsfs,  amssymb, dsfont, mathtools, slashbox, pifont, amsbsy, graphicx, url,  yfonts, bbm}
\usepackage[all,dvips,arc,curve,color,frame]{xy}

\newcommand\prefixtext[1]{%
  \ifvmode\else\\\@empty\fi
  \noalign{%
    \penalty0%
    \vbox{\mathstrut}%
    \penalty10000%
    \vskip-\baselineskip
    \penalty10000%
    \vbox to 0pt{%
      \normalbaselines
      \ifdim\linewidth=\columnwidth
      \else
        \parshape\@ne
        \@totalleftmargin\linewidth
      \fi
      \vss
      \noindent#1\par}%
      \penalty10000%
      \vskip-\baselineskip}%
      \penalty10000}

\DeclareMathAlphabet{\mathpzc}{OT1}{pzc}{m}{it}
\newcommand{\comment}[1]{}

\def\bep{\boldsymbol{\epsilon}}

\def\tr{\text{tr}}
\def\({\left(}
\def\){\right)}
\def\[{\left[}
\def\]{\right]}
\def\BEq{\begin{eqnarray}}
\def\EEq{\end{eqnarray}}
\def\BE*{\begin{eqnarray*}}
\def\EE*{\end{eqnarray*}}
\def\BA{\begin{array}}
\def\EA{\end{array}}

\def\0{\mathbf{0}}
\def\1{\mathbf{1}}
\def\a{\mathbf{a}}
\def\A{\mathbf{A}}

\def\C{\mathbf{C}}

\def\h{\mathbf{h}}
\def\H{\mathbf{H}}

\def\I{\mathbf{I}}

\def\M{\mathbf{M}}
\def\n{\mathbf{n}}
\def\N{\mathbf{N}}

\def\r{\mathbf{r}}
\def\R{\mathbf{R}}

\def\S{\mathbf{S}}

\def\x{\mathbf{x}}
\def\X{\mathbf{X}}

\def\Y{\mathbf{Y}}
\def\z{\mathbf{z}}
\def\Z{\mathbf{Z}}

\def\sc{\pmb{c}}

\def\and{\prefixtext{and}}

\def\diag{{\rm diag}}